\documentstyle[twoside,11pt,psfig]{article}
\def\deg{$^{\rm o}$}

\def\cc{\ifmmode {\rm cm}^{-3} \else cm$^{-3}$\fi}
\def\arcsec{\ifmmode '' \else $''$\fi}
\def\arcmin{\ifmmode ' \else $'$\fi}
\def\arcsecpoint{\ifmmode ''\!. \else $''\!.$\fi}
\def\arcminpoint{\ifmmode '\!. \else $'\!.$\fi}
\def\kms{\ifmmode {\rm km\ s}^{-1} \else km s$^{-1}$\fi}
\def\Hubble{\ifmmode {\rm km\ s}^{-1}\ {\rm Mpc}^{-1} 
	\else km s$^{-1}$ Mpc$^{-1}$\fi}
\def\ergsec{\ifmmode {\rm ergs\ s}^{-1} \else ergs s$^{-1}$\fi}
\def\ergscm2{\ifmmode {\rm ergs\ s}^{-1}\;{\rm cm}^{-2}
	  \else ergs s$^{-1}$ cm$^{-2}$\fi}
\def\ergscm2A{\ifmmode {\rm ergs\ s}^{-1}\;{\rm cm}^{-2}\;{\rm \AA}^{-1}
	  \else ergs s$^{-1}$ cm$^{-2}$ \AA$^{-1}$\fi}
\def\ergscm2Hz{\ifmmode {\rm ergs\ s}^{-1}\;{\rm cm}^{-2}\;{\rm Hz}^{-1}
	  \else ergs s$^{-1}$ cm$^{-2}$ Hz$^{-1}$\fi}
\def\Msun{\ifmmode {\rm M}_{\odot} \else M$_{\odot}$\fi}
\def\Lsun{\ifmmode {\rm L}_{\odot} \else L$_{\odot}$\fi}
\def\qo{\ifmmode q_{\rm o} \else $q_{\rm o}$\fi}
\def\Ho{\ifmmode H_{\rm o} \else $H_{\rm o}$\fi}
\def\ho{\ifmmode h_{\rm o} \else $h_{\rm o}$\fi}
\def\qo{\ifmmode q_{\rm o} \else $q_{\rm o}$\fi}
\def\ao{\ifmmode a_{\rm o} \else $a_{\rm o}$\fi}
\def\to{\ifmmode t_{\rm o} \else $t_{\rm o}$\fi}
\def\gtsim{\raisebox{-.5ex}{$\;\stackrel{>}{\sim}\;$}}
\def\vFWHM{\ifmmode v_{\mbox{\tiny FWHM}} \else
            $v_{\mbox{\tiny FWHM}}$\fi}
\def\Halpha{\ifmmode {\rm H}\alpha \else H$\alpha$\fi}
\def\Hbeta{\ifmmode {\rm H}\beta \else H$\beta$\fi}
\def\Hgamma{\ifmmode {\rm H}\gamma \else H$\gamma$\fi}
\def\Hdelta{\ifmmode {\rm H}\delta \else H$\delta$\fi}
\def\Lya{\ifmmode {\rm Ly}\alpha \else Ly$\alpha$\fi}
\def\Lyb{\ifmmode {\rm Ly}\beta \else Ly$\beta$\fi}
\def\hi{H\,{\sc i}}

\def\heii{He\,{\sc ii}}

\def\cii{C\,{\sc ii}}
\def\ciii{\ifmmode {\rm C}\,{\sc iii} \else C\,{\sc iii}\fi}
\def\civ{\ifmmode {\rm C}\,{\sc iv} \else C\,{\sc iv}\fi}

\def\niv{N\,{\sc iv}}
\def\nv{N\,{\sc v}}

\def\oiii{O\,{\sc iii}}
\def\oiv{O\,{\sc iv}}

\def\mgii{Mg\,{\sc ii}}
\def\siIV{Si\,{\sc iv}}
\def\siIII{Si\,{\sc iii}}

\def\aliii{Al\,{\sc iii}}
\def\o5007{[O\,{\sc iii}]\,$\lambda5007$}
\def\n{\footnotemark}
\begin{document}
\normalsize
\baselineskip 15pt
\begin{center}
{\bf MULTIWAVELENGTH OBSERVATIONS OF }\\
{\bf SHORT TIME-SCALE VARIABILITY IN NGC~4151.}\\
{\bf I.\ ULTRAVIOLET OBSERVATIONS}
\end{center}
\vskip 0.4cm
\small
\begin{center}
\sc
D.M.~Crenshaw,\n\
P.M.~Rodr\'{\i}guez-Pascual,\n\
S.V.~Penton,\n\
R.A.~Edelson,\n\
D.~Alloin,\n\
T.R.~Ayres,\footnotemark[3]\
J.~Clavel,\n\
K.~Horne,\n\
W.N.~Johnson,\n\
S.~Kaspi,\n\
K.T.~Korista,\n\
G.A.~Kriss,\n\
J.H.~Krolik,\footnotemark[11]\
M.A.~Malkan,\n\
D.~Maoz,\footnotemark[9]\
H.~Netzer,\footnotemark[9]\
P.T.~O'Brien,\n\
B.M.~Peterson,\n\
G.A.~Reichert,\n\
J.M.~Shull,\footnotemark[3]$^{,}$\n\
M.-H.~Ulrich,\n\
W.~Wamsteker,\footnotemark[2]\
R.S.~Warwick,\n\
T.~Yaqoob,\n\
T.J.~Balonek,\n\
P.~Barr,\footnotemark[6]\
G.E.~Bromage,\n\
M.~Carini,\n\
T.E.~Carone,\n\
F.-Z.~Cheng,\n\
K.K.~Chuvaev,\n\
M.~Dietrich,\n\
V.T.~Doroshenko,\n\
D.~Dultzin-Hacyan,\n\
A.V.~Filippenko,\n\
C.M.~Gaskell,\n\
I.S.~Glass,\n\
M.R.~Goad,\n\
J.~Hutchings,\n\
D.~Kazanas,\footnotemark[19]
W.~Kollatschny,\n\
A.P.~Koratkar,\footnotemark[32]
A.~Laor,\n\
K.~Leighly,\n\
V.M.~Lyutyi,\n\,
G.M.~MacAlpine,\n\
Yu.F.Malkov,\footnotemark[25]\
P.G.~Martin,\n\
B.~McCollum,\n\
N.I.~Merkulova,\footnotemark[25]\
L.~Metik,\footnotemark[25]\
V.G.~Metlov,\footnotemark[27]
H.R.~Miller,\n\
S.L.~Morris,\footnotemark[33]\
V.L.~Oknyanskij,\footnotemark[25]$^{,}$\n\
J.~Penfold,\n\
E.~P\'{e}rez,\n\
G.C.~Perola,\n\
G.~Pike,\footnotemark[19]\
R.W.~Pogge,\footnotemark[14]\
I.~Pronik,\footnotemark[25]\
V.I.~Pronik,\footnotemark[25]\
R.L.~Ptak,\n\
M.C.~Recondo-Gonz\'{a}lez,\footnotemark[2]\
J.M.~Rodr\'{\i}guez-Espinoza,\n\
E.L.~Rokaki,\n\
J.~Roland,\n\
A.C.~Sadun,\n\
I.~Salamanca,\footnotemark[5]\
M.~Santos-Lle\'{o},\footnotemark[5]\
S.G.~Sergeev,\footnotemark[25]\
S.M.~Smith,\footnotemark[14]\
M.A.J.~Snijders,\n\
L.S.~Sparke,\n\
G.M.~Stirpe,\n\
R.E.~Stoner,\footnotemark[46]\
W.-H.~Sun,\n\
E.~van~Groningen,\footnotemark[42]\
R.M.~Wagner,\footnotemark[14]$^{,}$\n\
S.~Wagner,\footnotemark[26]\
I.~Wanders,\footnotemark[14]\
W.F.~Welsh,\n\
R.J.~Weymann,\n\
B.J.~Wilkes,\n\
and W.~Zheng\footnotemark[11]
\end{center}
\vskip 0.4cm
\begin{center}
\it
Received\hspace{0.5em}\rule{2.0in}{0.01in}
\end{center}
\vskip 0.4cm
\rm
\leftskip=1.em
\parindent=-1.em

\setcounter{footnote}{0}

\n Computer Sciences Corporation, Laboratory for Astronomy and Solar 
Physics, NASA Goddard Space Flight Center, Code 681,
Greenbelt, MD  20771.

\n ESA IUE Observatory, P.O.\ Box 50727,
28080 Madrid, Spain.
 
\n Center for Astrophysics and Space Astronomy,
University of Colorado, Campus Box 389, Boulder, CO 80309.

\n Department of Physics and Astronomy, University of Iowa,
Iowa City, IA 52242.

\n Observatoire de Paris, URA 173 CNRS,
92195 Meudon, France.

\n ISO Observatory, Astrophysics Division of ESA,
ESTEC, Postbus 299, 2200-AG, The Netherlands.

\n School of Physics and Astronomy, University of St.\ Andrews,
North Haugh, St.\ Andrews KY16\,9SS, Scotland, United Kingdom.

\n Naval Research Laboratory, Code 4151, 4555 Overlook SW,
Washington, DC 20375-5320.

\n School of Physics and Astronomy and the Wise Observatory,
The Raymond and Beverly Sackler Faculty of Exact Sciences,
Tel-Aviv University, Tel-Aviv 69978, Israel.

\n  Department of Physics and Astronomy,
University of Kentucky, Lexington, KY  40506.

\n Department of Physics and Astronomy, The Johns Hopkins
University, Baltimore, MD 21218.

\n Department of Astronomy, University of California, 
Math-Science Building, Los Angeles, CA 90024.

\n Department of Astrophysics, Oxford University,
Keble Road, Oxford OX1 3RH, United Kingdom.

\n Department of Astronomy, The Ohio State University,
174 West 18th Avenue, Columbus, OH 43210.

\n Universities Space Research Association,
NASA Goddard Space Flight Center, Code 668,
Greenbelt, MD 20771.

\n Joint Institute for Laboratory Astrophysics;
University of Colorado and
National Institute of Standards and Technology, 
Campus Box 440, Boulder, CO 80309.

\n European Southern Observatory, Karl Schwarzschild Strasse 2,
85748 Garching, Germany.

\n Department of Astronomy, University of Leicester,
University Road, Leicester LE1\,7RH, United Kingdom 

\n Laboratory for High Energy Astrophysics, Code 665,
NASA Goddard Space Flight Center, Greenbelt, MD 20771.

\n Department of Physics and Astronomy, Colgate University,
Hamilton, NY 13346.

\n Centre for Astrophysics, University of Central Lancashire,
Preston PR1 2HE, U.K.

\n Center for Automated Space Studies, Department of Physics
and Astronomy, Western Kentucky University.

\n Space Sciences Laboratory, University of California,
Berkeley, CA 94720, and Eureka Scientific, Inc.

\n Center for Astrophysics, University of Science and Technology,
Hefei, Anhui, People's Republic of China.

\n Crimean Astrophysical Observatory, P/O Nauchny,
334413 Crimea, Ukraine.

\n Landessternwarte,  K\"{o}nigstuhl,
D-69117 Heidelberg, Germany.

\n Sternberg State Astronomical Institute, P/O Nauchny,
334413 Crimea, Ukraine.

\n Universidad  Nacional Autonoma de Mexico, Instituto de Astronomia,
Apartado Postal 70-264, 04510 Mexico D.F., Mexico.

\n Department of Astronomy, University of California,
Berkeley, CA  94720.

\n Department of Physics and Astronomy,
University of Nebraska, Lincoln, NE 68588.

\n South African Astronomical Observatory, P.O.\ Box 9,
Observatory 7935, South Africa.

\n Space Telescope Science Institute,
3700 San Martin Drive, Baltimore, MD 21218.

\n Dominion Astrophysical Observatory, 
5071 West Saanich Road, Victoria, B.C. V8X 4M6, Canada.

\n Universit\"{a}ts-Sternwarte G\"{o}ttingen,
Geismarlandstrasse 11, D-37083 G\"{o}ttingen, Germany.
 
\n Department of Astronomy, Caltech 130-33,
Pasadena, CA 91125.

\n Cosmic Radiation Laboratory, RIKEN, Hirosawa 2-1,
Wako, Saitama 351, Japan.

\n Sternberg Astronomical Institute, University of Moscow,
Universitetskij Prosp.\ 13, Moscow 119899, Russia.

\n Department of Astronomy, University of Michigan, Dennison Building, Ann
Arbor, MI  48109.

\n Canadian Institute for Theoretical Astrophysics,
University of Toronto, Toronto, ON M5S 1A1, Canada.

\n Computer Sciences Corporation, NASA Goddard Space Flight
Center, Code 684.9, Greenbelt, MD 20771.

\n Department of Physics and
Astronomy, Georgia State University, Atlanta, GA  30303.

\n Astronomiska observatoriet, Box 515, S-751 20 Uppsala, Sweden.

\n Department of Physics and Astronomy, University of Calgary,
2500 University Drive NW, Calgary, AB T2N 1N4, Canada, and
Department of Mathematics, Physics, and Engineering, 
Mount Royal College, Calgary T3E 6K6, Canada.

\n Instituto de Astrof\'{\i}sica de Andalucia,
Aptdo.\ 3004, 18080 Granada, Spain.

\n Istituto Astronomico dell'Universit\`{a}, Via Lancisi 29, I-00161 Rome,
Italy.

\n Department of Physics and Astronomy, Bowling Green State University, 
Bowling Green, OH  43403.

\n Instituto de Astrof\'{\i}sica de Canarias, E-38200 La Laguna, Tenerife,
Spain.

\n Royal Observatory Edinburgh, University of Edinburgh, Blackford Hill,
Edinburgh EH9\,3HJ, United Kingdom.

\n Institut d'Astrophysique, 98 bis Boulevard Arago,
F-75014 Paris, France.

\n Department of Physics and Astronomy and Bradley Observatory, Agnes Scott
College, Decatur, GA 30030.

\n IRAM, 300 Rue de la Piscine, 38046 Saint Martin d'Heres,
France.

\n Department of Astronomy, University of Wisconsin, 475 N.\ Charter Street,
Madison, WI  53706.

\n Osservatorio Astronomico di Bologna, Via Zamboni 33, I-40126,
Bologna, Italy.

\n Institute of Astronomy, National Central University, 
Chung-Li, Taiwan 32054, Republic of China.

\n Mailing address: Lowell Observatory, Mars Hill Road,
1400 West, Flagstaff, AZ  86001.

\n Department of Physics, Keele University, Keele ST5 5BG,
Staffordshire, United Kingdom.

\n Observatories of the Carnegie Institution of Washington,
813 Santa Barbara Street, Pasadena, CA  91101.

\n Harvard-Smithsonian Center for Astrophysics, 
60 Garden Street, Cambridge, MA  02138.

%
%

\newpage
\normalsize
\baselineskip 24pt
\parindent=0.2in
\begin{center}
{\bf ABSTRACT}
\end{center}

\medskip

We present the results of an intensive ultraviolet monitoring campaign
on the Seyfert 1 galaxy NGC 4151, as part of an effort to study its
short time-scale variability over a broad range in wavelength. The
nucleus of NGC 4151 was observed continuously with the {\it
International Ultraviolet Explorer} (IUE) for 9.3 days, yielding a
pair of LWP and SWP spectra every $\sim$70 minutes, and during
four-hour periods for 4 days prior to and 5 days after the continuous
monitoring period. The sampling frequency of the observations is an
order of magnitude higher than that of any previous UV monitoring
campaign on a Seyfert galaxy. 

The continuum fluxes in bands from 1275 \AA\ to 2688 \AA\ went through
four significant and well-defined ``events'' of duration 2 -- 3 days
during the continuous monitoring period. We find that the amplitudes
of the continuum variations decrease with increasing wavelength, which
extends a general trend for this and other Seyfert galaxies to smaller
time scales (i.e., a few days). The continuum variations in all of the
UV bands are {\it simultaneous} to within an accuracy of about 0.15
days, providing a strict constraint on continuum models. The
emission-line light curves show only one major event during the
continuous monitoring (a slow rise followed by a shallow dip), and do
not correlate well with continuum light curves over the (short)
duration of the campaign, because the time scale for continuum
variations is apparently smaller than the response times of the
emission lines. 

\medskip

\noindent
{\it Subject headings:} galaxies: individual (NGC 4151) -- 
galaxies:active -- galaxies:Seyfert -- ultraviolet:spectra

\vfill\eject

\begin{center}
{\sc 1. Introduction}
\end{center}

Variability monitoring of active galactic nuclei (AGN) has become a 
very productive way to probe the spatially unresolved nuclear
continuum source and, when present, surrounding broad-line region
(BLR). The success of recent large-scale monitoring campaigns are due
to high temporal sampling rates over extended periods of time (see
Peterson 1993 for a review). The cornerstone of most of these
campaigns has been the International Ultraviolet Explorer (IUE),
because it can provide long periods of observations at precise
intervals, accurate absolute flux levels, and access to the UV, where
the continuum and high-ionization lines are more strongly variable
than in the optical. Most campaigns have focused on nearby bright 
Seyfert 1 galaxies whose UV continua and emission lines were previously 
known to be strongly variable.

The initial large IUE campaign on NGC 5548 is described by Clavel et
al. (1991), and results from concurrent and subsequent ground-based
monitoring programs are given in Peterson et al. (1991, 1992, 1994),
and Dietrich et al. (1993). One of the most fundamental results from
these efforts is that there was no detectable delay between the
variations in the ultraviolet continuum bands and those in the
optical; that is, the time lag between the UV and optical light curves
was $\leq$ 4 days. This provides an important constraint on models of
the continuum source. For example, for thin accretion disks (e.g.,
Shakura \& Sunyaev 1973), this implies that surprisingly high radial
signal speeds ($\gtsim 0.1 c$) coordinate the different regions of the
disk (Krolik et al. 1991). A possible explanation is that the UV and
optical continuum emission is due to reprocessing by cooler, outer
material of X-ray photons created closer in (Courvoisier \& Clavel
1991; Collin-Souffrin 1991; Krolik et al. 1991). 

A major campaign on NGC 3783 with IUE (Reichert et al. 1994) and
ground-based telescopes (Stirpe et al. 1994) resulted in the same
approximate upper limit ($\leq$4 days) for the lag between optical and
UV continuum variations. A subsequent HST, IUE, and ground-based
campaign on NGC 5548 (Korista et al. 1995), anchored by daily
observations with the Faint Object Spectrograph, demonstrated that the
UV and optical continuum variations in NGC 5548 were further
constrained to be simultaneous to within $\pm$1 day. In addition,
Clavel et al. (1992) show that the X-ray and UV continuum fluxes in
NGC 5548 are correlated, but with considerable scatter and a rather
loose constraint of $\leq$ 6 days on the time lag. It has become
evident that multiwavelength monitoring projects with even higher
temporal resolution are needed, in order to obtain tighter constraints
on the lags, if any, between X-ray, UV, and optical continuum
variations. 

The previous campaigns have also demonstrated that the emission-line
response times to changes in the photoionizing continuum are very
short (days) and a function of ionization, with the high
ionization lines responding more rapidly. In fact, the initial
campaigns on NGC 5548 and NGC 3783 (Clavel et al. 1991; Reichert et
al. 1994) found that the lags for the highest ionization lines, \heii\
$\lambda$1640 and \nv\ $\lambda$1240, were $\leq$ 4
days. With the higher sampling of the subsequent HST, IUE, and
ground-based campaign on NGC 5548, Korista et al. (1995) were able to
determine that the lags for these lines were slightly less than 2
days. Thus, another reason for obtaining higher temporal resolution
is to check this result for this and other Seyferts, and specifically
to fully resolve the transfer function of the high ionization lines
(Peterson 1993). 

We report on a new campaign to provide an order of magnitude increase
in the sampling rate over previous campaigns on Seyfert 1 galaxies,
similar to that obtained for the BL Lac object PKS 2155--304,
which was monitored continuously by IUE for 5 days (Urry et al. 1993)
as part of a multiwavelength campaign (Edelson et al. 1995). We also 
made a determined effort to obtain concurrent observations of
NGC 4151 at other wavelengths, particularly in the optical and X-ray
regions, to test the predictions of accretion disk and continuum
reprocessing models. This would also allow a comparison with the
multiwavelength observations of PKS 2155-304, an object with a strong
{\it beamed} component. The data and basic results from the IUE
campaign on NGC 4151 are given in this paper. Other papers in this
series report on optical observations (Kaspi et al. 1996, Paper II),
high-energy observations (Warwick et al. 1996, Paper III), and a
comparison of the multiwavelength continuum data (Edelson et al.
1996, Paper IV). 

NGC 4151 is a nearby (cz = 995 \kms) barred spiral galaxy that is
viewed nearly face-on (Simkin 1975). It was classified as a Seyfert
1.5 by Osterbrock \& Koski (1976), because its nucleus shows strong
narrow components for the permitted lines, in addition to the broad
(thousands of \kms\ FHWM) permitted and narrow (hundreds of \kms\ FWHM)
forbidden lines that define a Seyfert 1 galaxy. HST images show that the
narrow-line [\oiii] $\lambda$5007 emission arises from a nuclear point
source and an extended ($\sim$ 3\arcsec) NLR that consists of a number
of emission-line clouds in a biconical structure (Evans et
al.\ 1993). The radio emission is extended along the same general
direction as the [\oiii] emission on arcsecond and sub-arcsecond
scales (Johnston et al. 1982; Wilson \& Ulvestad 1983), although the
optical emission-line and radio axes are misaligned by $\sim$ 20\deg.
NGC 4151 exhibits a complex X-ray spectrum, which can be characterized
in the 2 - 10 keV region by a power-law continuum modified by a warm
or partial absorber and a soft X-ray excess in the 0.1 -
2 keV range (Holt et al. 1980; Yaqoob, Warwick, \& Pounds 1989; 
Weaver et al. 1994a,b).

Because it is so bright and strongly variable in the UV, NGC 4151 is
the ideal target for intensive monitoring (see Ulrich et al. 1991 for
a summary of previous UV observations). It shows ultraviolet continuum
variations with doubling times as small as a week (Clavel et al.
1990), and is one of the small number of AGN for which emission-line
lags have been reliably determined by cross-correlation of the
emission and continuum light curves. The most accurate lags determined
for NGC 4151 thus far are $4 \pm 3$ days for \civ\ $\lambda$1549
(Clavel et al. 1990) and $9 \pm 2$ days for the Balmer lines (Maoz et
al. 1991), confirming earlier estimates by Gaskell \& Sparke (1986)
and Peterson \& Cota (1988). The UV spectrum of NGC~4151 shows
extremely broad emission lines ($\sim$30,000 \kms\ FWZI for \civ). It
also contains a number of broad (1000 \kms), blue shifted ($-$1100 to
$-$100 \kms), and variable absorption lines that arise in ions of
widely different stages (Bromage et al. 1985; Kriss et al. 1992), and
two unidentified emission lines, known as L1 $\lambda$1518 and L2
$\lambda$1594, that bracket the \civ\ $\lambda$1549 feature (Ulrich et
al.\ 1985; Clavel et al. 1987).

\bigskip

\begin{center}
{\sc 2. Observations}
\end{center}

The nucleus of NGC 4151 was observed with the IUE SWP (1150 -- 1970
\AA) and LWP (1970 -- 3300 \AA) cameras through the large apertures
(10\arcsec x 20\arcsec) in low-dispersion mode (resolution $=$ 5 -- 8
\AA\ FWHM). Observations were made in a continuous mode over 9.3 days
during 1993 December 1 -- 10. In addition, observations were obtained
during four-hour US2 shifts (which frequently experience higher
particle radiation) on the four days prior to and five days after the
continuous monitoring period. The standard observing procedure was to
obtain alternate LWP and SWP exposures by reading and preparing one
camera while the other camera was exposing, which resulted in a pair
of spectra every $\sim$70 minutes. Typical exposure times were 20 min
for the SWP images and 10 min for the LWP images. During each day of
the continuous monitoring, the observations were interupted for
$\sim$2 hours as the Earth occulted the target and the spacecraft was
maneuvered to a low $\beta$ (angle between the telescope axis and the
anti-solar direction) to maintain attitude control and cool the
onboard computer. 

The observations were affected by the presence of scattered solar (and
occasionally Earth) light in the telescope tube, which has been
present since early 1991 and is strong at $\beta$ $\geq$ 50\deg
(Carini \& Weinstein 1992). In order to obtain concurrent observations
with other satellites (e.g., ROSAT) it was necessary to observe NGC
4151 at $\beta$ $\approx$ 90\deg. The scattered light spectrum is such
that there is contamination of the LWP spectra at the long-wavelength
end (see section 3.2), but no contamination of the SWP spectra. The
most noticeable effect of the scattered light is that it greatly
increases the background level in the FES, which is the optical target
acquisition detector. Thus, the nucleus of NGC~4151 could not be
detected directly, since the FES background counts exceeded those
expected for the target by a factor of $\sim$50, and no optical light
curve could be obtained from the FES. Fortunately, the scattered light
had little effect on acquisition and guiding during the exposures. The
nucleus of NGC~4151 was acquired by blind offset from a nearby bright
star (SAO 62869), which is a procedure that typically results in a
positioning error in the aperture that is $<$ 1\arcsec. During the
exposures, the same bright star was used for guiding, since it
remained in the portion of the FES field-of-view that is least
affected by the scattered light. The offset slew was repeated about
once every 8 hours to recenter the target in the aperture and to
update the guide star position, since the spacecraft rolls to maintain
optimal positioning of the solar arrays. 

The details of individual observations can be obtained from the IUE
Merged Log, which is available through the IUE data analysis centers
at NASA Goddard Space Flight Center and the ESA Villafranca Satellite
Tracking Station (VILSPA); the progam ID's for these observations are
``IMPRE'' (NASA) and ``QQ105'' (VILSPA). For example, there was
significant particle radiation for some of the images obtained during
the US2 shifts, which could result in slightly lower signal-to-noise
ratios for the affected spectra. These spectra can be identified in
the merged log from background levels that are higher than the
average background levels, which are typically $\sim$15 DN (data
numbers) for the SWP and $\sim$27 DN for the LWP. 

A total of 205 SWP and 196 LWP spectra were obtained of NGC 4151
during the campaign. Only six images are considered to be unusable.
The exposure time for SWP 49394 was cut short due to an impending
Earth occultation. The target was on the edge of the aperture for LWP
26984, SWP 49428, and LWP 26895. Most of the two images for LWP 26931
and LWP 27008 were lost due to telemetry problems, and could not be
recovered. The other problems encountered are minor, and do not
significantly affect the measured fluxes. We are left with 395 useful
spectra to work with: 203 SWP and 192 LWP. 

\bigskip

\begin{center}
{\sc 3. Data Reduction and Analysis}
\end{center}

The IUE project has developed techniques for improving the
signal-to-noise, wavelength assignment, and flux calibration of IUE
spectra; these techniques are being used in the new processing system
(NEWSIPS) to produce the IUE Final Archives. However, at the time of
the observations, only the old processing system (IUESIPS) was
available for current data. We decided to use a newly available system
developed by Tom Ayres called ``TOMSIPS''.

\bigskip

\begin{center}
{\it 3.1 TOMSIPS Reduction}
\end{center}

TOMSIPS is based on many of the techniques developed for NEWSIPS and
includes a realistic noise model (Ayres 1993). TOMSIPS, like NEWSIPS,
uses an identically rotated intensity transfer function (ITF). The old
IUESIPS used pre-rotated ITFs, which do not always match up with the
current image and can introduce fixed pattern noise. TOMSIPS uses an
ITF based directly on the raw images of the flux standard white dwarf
G191B2B, and a wavelength calibration based upon the emission-line
spectra of $\lambda$ Andromeda. 

TOMSIPS uses a slit-weighted extraction method similar to the OPTIMAL
techique (Kinney, Bohlin, \& Neill 1991), with the distinct
difference that the cross-dispersion profile is not of a fixed form,
but matches the actual average cross-dispersion profile in the region.
Because the cross-dispersion profile is both wavelength and
emission-line dependent, 10 separate cross-dispersion regions are used
for the SWP and 7 are used for the LWP. Unlike previous extraction
techniques, including OPTIMAL and GEX (Gaussian extraction, see
Reichert et al. 1994), the TOMSIPS errors are not ``extraction''
errors, but are true estimates of the flux uncertainties empirically
derived from an independent noise model (Ayres 1993).  The improved
noise model and similarly processed ITF substantially reduces the
pixel-to-pixel variations of the final spectra. A detailed comparison
of TOMSIPS with other processing techniques appears in Penton et al.
(1996). 

\bigskip

\begin{center}
{\it 3.2 Scattered Light Contamination of LWP Spectra}
\end{center}

The scattered light in the IUE telescope is characterized by a solar
spectrum (Carini \& Weinstein 1992), and therefore rises sharply at
the long-wavelength end of the LWP region. Unfortunately, the
scattered light exhibits a strong (and possibly variable) gradient
across the aperture, and there are no proven techniques for removing
it at this time. The flux levels of several sky background LWP spectra
obtained during the campaign indicate that the scattered light
contributed the following approximate percentages to the total fluxes
in the continuum bands: $\sim$1\% at 2300 \AA, $\sim7$\% at 2688 \AA,
$\sim$26\% at 2970 \AA, and $\sim$37\% at 3130 \AA. Checks of the flux
levels inside the aperture, but away from the spectra, indicate that
the background contribution varied by $\leq$1\% at 2688 \AA\ over the
course of the campaign. At longer wavelengths, the background
variation started to significantly alter the observed continuum
variations. 

\bigskip

\begin{center}
{\it 3.3 Continuum and Line Measurements}
\end{center}

Continuum measurements are made in known line-free bands in the rest
frame of NGC 4151. A continuum flux is taken to be the error-weighted
mean over the bandpass, and the continuum flux error is the standard
deviation of the mean over the bandpass. For SWP spectra, the
continuum regions selected are 1260 -- 1290~\AA, 1420 -- 1460~\AA, and
1805 -- 1835~\AA. In the LWP spectra, the only usable continuum band
is 2625 -- 2750~\AA; at shorter wavelengths, the spectra are too
noisy, and at longer wavelengths, the spectra are too contaminated by
the variable scattered light. The continuum bands are shown in Figure
1, along with the average SWP and LWP spectra for this campaign. The 
sharp upturn at the long-wavelength end is due to the scattered light 
contamination.

To measure the emission and absorption line components, pre-defined
regions around each feature of interest are extracted and a fit is
made in the rest frame. Initially, the fit consists of a power-law
continuum of the form $F_{\lambda}=F_{0} ({\lambda /
\lambda_{0}})^\alpha $, with the data weighted by the TOMSIPS
errors. Gaussian components of the form
\begin{equation}
 F^{G}_{\lambda}\,  =\,  F_0\, \exp\left[- \frac{(\lambda
-\lambda_c)^2}{2\sigma_G^2}\right] 
\end{equation}
are added to measure the emission and absorption features.  The
Gaussians are initially centered at the expected line center
($\lambda_c$), but are allowed to ``float'' in wavelength ($\lambda$),
width ($\sigma_G$), and amplitude ($F_0$). The float in wavelength and
width is required to compensate for wavelength calibration errors and
the asymmetry of the broad emission lines. All Gaussian components
without an initially fixed minimum width are constrained to have a
minimum width of twice the instrumental profile ($\sigma_G$ $=$
1.5~\AA) to prevent fitting spurious pixels. Five separate regions
containing emission and/or absorption features were examined
separately, since slightly different procedures are needed to properly
fit each spectral feature. These regions are: 1155 -- 1315~\AA\
(Ly$\alpha$ and \nv\ $\lambda$1240), 1307 -- 1463~\AA\ (\cii\
$\lambda$1334, \siIV\ $\lambda$1398 and \oiv] $\lambda$1402), 1425 --
1750~\AA\ (\niv] $\lambda$1486, \civ\ $\lambda$1549, \heii\
$\lambda$1640, and \oiii] $\lambda$1663), 1790 -- 1960~\AA\ (\aliii\
$\lambda$1857, \siIII] $\lambda$1890, and \ciii] $\lambda$1909), and
2640 -- 2915~\AA\ (\mgii\ $\lambda$2800). 

To ensure an unbiased extraction of the parameters characterizing the
UV emission and absorption features, the spectral components are
systematically fit with a modified version of the MINUIT (James
\& Roos 1975) software. The CERN-developed MINUIT uses the non-linear
least squares Levenberg-Marquart method to fit $\chi^2$  minimizing
Gaussian components to the spectral features. Errors in the
fitting parameters are determined by exploring parameter space near
the minimum and are reported as 1$\sigma$ errors.

Complicated features such as \civ\ require multiple Gaussian
components. While each new component will reduce $\chi^{2}$, it may
not always be statistically significant. In these cases, an F-test is
applied (Bevington 1969). Only when the component passes the F-test
will it be added to the final fit. As an additional restriction, we
have chosen to limit the maximum number of components for each specific
line (e.g., \civ\ $\lambda$1549) to three. All Gaussians were initially
allowed a wide range of parameter space to achieve the best fit; this
range was reduced as obvious trends of the fits became apparent.

Line fluxes were calculated by integrating and adding together the
individual components of the feature of interest. A quadrature sum of
the individual integrated Gaussian errors is not a true error estimate
of the integrated flux, but in fact overestimates the true error.
Similar to a method used in a previous campaign on NGC~5548 (Clavel et
al. 1991), the point-to-point integrated light curve variations were
used to scale the errors to the proper values.

Table 1 shows the allowed ranges and means of the final component
fits. The components that were summed together to produce a
measurement for a particular line or blend are described in the next
subsection. Figure 2 shows an example of the combined fit to an
individual SWP spectrum. 

\bigskip

\begin{center}
{\it 3.4 Summation of Line Components}
\end{center}

We attach no physical signficance to individual components for a
particular line (e.g., \civ\ $\lambda$1549); the component fits are
just a convenient description of the data. In addition, many of the
individual emission and absorption lines in a given region are blended
as a result of their proximity, and their individual light curves are
noisy and difficult to interpret as a result of the fitting
technique's inability to accurately deconvolve them. Therefore, as in
the past (Clavel et al. 1991; Reichert et al. 1994), we use the sum of
components to represent the dominant emission-line in a particular
region. When we quote results for a particular sum of components, we 
use the dominant emission feature as a designation (e.g., ``\civ'' for 
the sum of the components of \civ\ and \niv]).

Many of the features in NGC 4151 are very difficult to measure
accurately due to contamination and/or the complicated nature of the
UV spectrum (many broad and narrow emission and absorption lines).
Their light curves at relatively low levels of variability are 
very noisy, and cannot be used for the detailed analyses in Section 4.
The redshift of NGC~4151 places its Ly$\alpha$ emission at 1219.7~\AA,
which is too close to the geocoronal emission to allow a separate fit
to the two narrow components. Hence, the Ly$\alpha$ $+$ \nv\ light curve
is dominated by the substantial variation of geocoronal Ly$\alpha$
over the course of each day, and is not usable for this study. 
As shown in Table 1, the \siIV\ $+$ \oiv] feature is dominated by
a strong \siIV\ absorption doublet, and an accurate measurement of the 
intrinsic emission is not possible. Finally, the \mgii\ feature is 
strongly affected by the scattered light, particularly in the red 
wing.

The fluxes of the features around \civ, \heii, and \ciii] can be 
measured accurately enough to produce reasonably good light curves.
The \civ\ line profile is complicated, showing multicomponent
emission and self-absorption. Due to their proximity, the \civ, \niv],
\heii, and \oiii] features are all fit simultaneously. The \niv],
\oiii], and \heii\ features are each fit with a single Gaussian, while
the more complicated \civ\ feature is fit with three Gaussian
components: a narrow emission component, a very narrow absorption
component, and an extended emission component. The \heii\ and \oiii]
emission are blended, but distinct enough from the \civ\ emission to
be treated as a separate feature. Thus, we have separate measurments
for \civ\ $+$ \niv] and \heii\ $+$ \oiii].

The \ciii] region is modeled with an absorption component for \aliii,
a single emission component for \siIII], and up to three emission
components for \ciii], as shown in Table 1. Due to the asymmetry of
the \ciii] emission, three components are needed: a narrow component,
a broad component, and a component labeled ``red''. The red component
is limited in central wavelength to avoid interference with the \siIII]
emission on the blue wing of \ciii]. The \aliii\ absorption on the
extreme blue wing of the emission can be separated from the overall
feature, so the measurement is for \ciii] $+$ \siIII]. 

\bigskip

\begin{center}
{\it 3.5 Comparison with IUESIPS}
\end{center}

As a consistency check on the TOMSIPS processing scheme, we compared the
measured continuum fluxes with those obtained in the same wavelength
bins from the IUESIPS spectra. Figure 3 show this comparison for the
1275 \AA\ bin, demonstrating that the fluxes from the two methods are
extremely well correlated (the linear correlation coefficient is r $=$
0.96). The other SWP continuum bins show the same excellent
correlation, with the TOMSIPS fluxes systematically higher than the
IUESIPS fluxes by 1 -- 10\%, depending on the bin. This is a direct
consequence of the slightly different (and improved) photometric
correction and absolute sensitivity calibration as a function of
wavelength.

\bigskip

\begin{center}
{\it 3.6 The Effects of Reddening}
\end{center}

In principle, the effects of reddening by dust on the intrinsic
continuum and emission-lines fluxes can be very significant. However,
we note that our principal results are on the fractional amplitudes
and time scales of the variations, which are not affected by
reddening. In addition, a variety of estimates indicate that reddening
is quite weak in NGC 4151. The Galactic reddening is essentially zero:
the \hi\ column density of 2.1 x 10$^{20}$ cm$^{-2}$ (Stark et al.
1992) gives E$_{B-V}$ $\leq$ 0.01 (according to Burstein and Heiles
1982). Penston et al. (1991) show that the {\it total} reddening of
the UV continuum determined from the 2200 \AA\ feature is E$_{B-V}$
$\leq$ 0.1; we have verified this result using the average NGC 4151
spectrum for this campaign and by assuming the interstellar extinction
curve of Savage and Mathis (1979) is applicable. Kriss et al. (1992)
calculate E$_{B-V}$ $=$ 0.039 from a power-law plus Galactic
extinction fit to the far-UV spectrum of NGC 4151. For the narrow
emission-lines, Kriss et al. (1992) determine that E$_{B-V}$ $\leq$
0.12 from the recombination lines \heii\ $\lambda$1085 and \heii\
$\lambda$1640. There is currently no accurate way to determine the
reddening of the broad components of the lines, although it is not
expected that dust can survive in the broad-line region of NGC 4151
(Ferland and Mushotzky 1982). 

\bigskip

\begin{center}
{\sc 4. Results}
\end{center}

Tables 2 and 3 give the TOMSIPS continuum and line fluxes (and
associated errors) as a function of Julian Date at the {\it midpoint}
of the SWP and LWP exposures. The image numbers are given so that data
points can be be identified with specific images. The few observations
of NGC 4151 that are not included in these tables are listed in
Section 2. 

\bigskip

\begin{center}
{\it 4.1 Pattern of Variability: Continuum Bands}
\end{center}

Figure 4 gives the SWP and LWP continuum light curves as a function of
Julian Date. The light curves show significant variations on a number
of different time scales, particularly at short wavelengths. The
variations are as large as 40 -- 50\% on a time scale of several days,
and $\sim$10\% on a time scale of several hours. During the 9.3 days of
continuous monitoring, there were four large-amplitude ``events'' of
duration 2 -- 3 days (minimum to minimum). These events are temporally
well resolved, and they are easily recognized in each continuum
waveband. Many of the shorter time-scale, small-amplitude features
also repeat in at least two different wavebands. The variations prior
to and after the continuous monitoring period are clearly undersampled
and are of limited use, although it is evident that there were strong
variations during the last five days of the monitoring period. 
It is clear from inspection of Figure 4 that the continuum variations
are all highly correlated, with no discernable lag between the light
curves. Paper IV will show that the continuum variations in different
UV bands are all simultaneous to within $\pm$ 0.15 days. 

Table 4 lists some basic properties of the variability in each
waveband. The mean fluxes for the entire data set are given, as well
as the mean errors, which are the average values of flux error divided
by mean flux. F$_{var}$, the fractional variability, is the standard
deviation of the fluxes divided by the mean flux in each waveband. It
has been corrected to reflect the intrinsic variability by subtracting
the mean error in quadrature. R$_{max}$ is the ratio of largest to
smallest mean flux in each waveband. It is clear from the parameters
in Table 4 and from inspection of the light curves that there were
significant variations in all continuum wavebands over this relatively
short period of time. In particular, F$_{var}$ is 4 -- 10 times larger
than the mean error, which is only about 1 -- 1.5\%. 

The amplitude of the continuum variations decreases with increasing
wavelength, as was the case for NGC 5548 (Clavel et al. 1989) and NGC
3783 (Reichert et al. 1994). This can be seen in both the F$_{var}$
and R$_{max}$ parameters. This result is in general agreement with
previous IUE studies of NGC 4151 (e.g., Perola et al. 1982), which
find that for larger amplitude (but undersampled) variations, the UV
continuum radiation hardens as it brightens. The combination of the
well-sampled UV variations described in this paper with those in the
optical (Paper II) and X-ray regions (Paper III) permits a more
detailed examination of the behavior of continuum amplitude as a
function of wavelength (see Paper IV). 

\bigskip
\newpage

\begin{center}
{\it 4.2 Pattern of Variability: Emission Lines}
\end{center}

Figure 5 gives the light curves for the strongest lines in the SWP
region (as well as the continuum light curve at 1275 \AA\ for
comparison). The emission-line light curves are similar in appearance,
with the fluxes rising through the first half of the campaign (over 8
-- 9 days), and leveling off thereafter. \civ, the line with the
smallest percentage errors, shows evidence for a subsequent shallow
dip of duration $\sim$3 days in its light curve before recovering to
the previous maximum. The \heii\ and \ciii] light curves appear to
reflect these trends, although they are noisier. There are also more
rapid variations in the emission lines, particularly in the case of
\heii, which shows 15 -- 20\% variations over a time scale of $\sim$ 1
day. We cannot identify any source of systematic error that might
produce these variations, and therefore assume that they are real. 

The light curves of the emission-lines are substantially different in
character than the continuum light curves. The large-amplitude
continuum variations are very well defined and much more rapid than
the emission-line variations. Presumably, the difference in the
light curves is the result of a substantial response time of the lines
to changes in the continuum, which will be explored in the next
subsection.

Some basic properties of the emission-line variations are listed in
Table 4. The fractional variability F$_{var}$, again corrected to
reflect the intrinsic variability, and the ratio of largest to
smallest flux R$_{max}$ show that there were significant variations
for all three emission lines. There is no obvious trend of larger
variability amplitude for higher ionization lines, as was the case for
NGC 5548 (Clavel et al. 1991) and NGC 3783 (Reichert et al. 1994),
although we are restricted to only a few lines due to the small
amplitude of variations over a short period of time and the
complicated nature of the spectrum of NGC 4151. In light of previous
studies, the larger amplitude of the \heii\ variations is expected,
but it is somewhat surprising that the amplitude of the \ciii]
variations is larger than that of \civ. 

\bigskip

\begin{center}
{\it 4.3 Time-Series Analysis}
\end{center}

Cross-correlation of a continuum light curve with a light curve from
another continuum band or an emission-line light curve has been used
in the past to determinine if the variations are correlated and if
there is a time lag between the two series. Cross-correlations between
the UV continuum light curves are given in Paper IV, to facilitate
comparisons with other wavelength regions. For this paper, the
continuum light curve at 1275 \AA\ was cross-correlated with that of
each emission-line feature, and in addition, the \civ\ light
curve was cross-correlated with itself to generate its
auto-correlation function (ACF). Two distinct correlation functions
were calculated: the interpolation cross-correlation function (CCF;
cf. Gaskell \& Sparke 1986; Gaskell \& Peterson 1987) and the discrete
correlation function (DCF; cf. Edelson \& Krolik 1988). The
correlations were calculated in the manner described by White \&
Peterson (1994). The calculations were performed for the subset of data
obtained during the continuous monitoring period, and for all of the
data obtained during the IUE campaign. The sampling interval chosen
was 0.05 days (72 min), which is the approximate interval between
consecutive observations with a particular camera during the
continuous monitoring period. 

Figures 6 and 7 show the emission-line CCF's and DCF's for the
continuous and entire data sets, respectively. The uniform sampling is
responsible for the agreement between CCF's and DCF's in Figure 6. In
Figure 7, after a longer span of data is included, both CCF and DCF
values are higher in the 1 -- 5 day regime, and additional peaks in
the correlation functions appear. The CCF and DCF values for the 
entire data set are similar, but do not agree exactly because the 
weighting is different; the DCF is based on the data points only, 
whereas the CCF relies on interpolated data as well (see Gaskell, 
Koratkar, \& Sparke 1988).

Figure 6 indicates that the emission-line variations are not well
correlated with the continuum variations; the peak values of the CCF's
are only 0.42, 0.58, and 0.47 for \civ, \heii, and \ciii]
respectively. This is not a surprise, since the continuum and
emission-line light curves are so different. In addition, the CCF's
and DCF's shown in Figure 7 for the entire data set exhibit multiple
peaks. This is a result of the fact that the continuum light curves
show several quick events during the monitoring period, whereas the
emission-line light curves show only one well-defined event of longer
duration. The best case for a significant result comes from the \heii\
correlations, which show a fairly-well defined peak at a lag of
$\sim$0.2 days. This {\it suggests} the presence of a component of
\heii\ emission that responds to continuum variations on very short
time scales, although this component does not appear to be the major
one. 

The principal time scale for continuum changes in this campaign
appears to be substantially shorter than the characteristic response
times of the emission-lines, as indicated by the small FWHM of the ACF
for the 1275 \AA\ continuum band (1.2 days, see Paper IV) compared to
that for \civ\ (4.4 days, see Figure 6). Since the continuum and
emission-line light curves are so dissimilar, the correlation
functions are of limited use, and the lags at which the
cross-correlations peak are not tabulated here. A more realistic value
for the \civ\ emission-line lag has been determined with a data set of
longer duration by Clavel et al. (1990), as discussed in the
Introduction. 

\bigskip

\begin{center}
{\sc 5. Summary and Conclusions}
\end{center}

We have observed the nucleus of NGC 4151 with IUE continuously for 9.3
days, obtaining a pair of SWP and LWP spectra every $\sim$70 min, in
the most intensive UV monitoring campaign to date for a Seyfert 1
galaxy. Observations were also obtained on the four days prior to and
five days after the continuous monitoring period. The IUE observations
are part of a multiwavelength effort to study the short time-scale
(hours to days) variations of NGC 4151, which have not been well
characterized in the past for any Seyfert 1 galaxy. 

During the monitoring period, significant variations were detected in
the fluxes of the continuum bands and the emission-line features. For
the continuous monitoring period, there are four well-defined
``events'' in the UV continuum light curves, whereas the light curves
for the strong emission lines are very different, primarily showing a
a slow rise followed by a shallow dip. Measurement and
cross-correlation of the light curves allow us to draw some important
conclusions: 

1. The UV continuum of NGC 4151 can vary significantly on very short 
time scales, going through an ``event'' (i.e., a significant local 
maximum preceded and followed by local minima) in only 2 -- 3 days.
The amplitudes of the events in this case are small compared to those 
found in NGC 4151 over longer time scales (Clavel et al. 1990), but 
are large compared to the IUE errors, demonstrating the feasibility 
and importance of continuous monitoring of AGN in the UV.

2. The relative amplitudes of the continuum variations decrease with
increasing wavelength: the ratios of largest to smallest flux value
over the 18-day monitoring period are R$_{max}$ $=$ 1.51, 1.45, 1.31,
and 1.24 at $\lambda$ $=$ 1275 \AA, 1440 \AA, 1820 \AA, and 2688 \AA.
This behavior has also been seen in the monitoring campaigns on NGC
5548 (Clavel et al. 1991; Korista et al. 1995) and NGC 3783 (Reichert
et al. 1994) on longer time scales. 

3. The continuum variations in all of the UV bands are {\it
simultaneous} to within $\pm$0.15 days (see Paper IV). This is an
important and very strict constraint compared to the upper limits on
UV continuum lags obtained for NGC 5548 ($\Delta$t$_{peak}$ $\leq$ 4
days, Clavel et al. 1991; $\Delta$t$_{peak}$ $\leq$ 1 day, Korista et
al. 1995) and NGC 3783 ($\Delta$t$_{peak}$ $\leq$ 4 days, Reichert et
al. 1994). 

4. The emission-line variations of NGC 4151 are not always well
correlated with the continuum variations over short periods of time
(days or less). The apparent reason for the dissimilar continuum and
emission-line light curves from our observations is the relatively
short time scale for continuum variations compared to the response
times of the emission-lines. Consequently, cross-correlations are not
very useful tools in this case; better tools (e.g., techniques for 
determining the transfer function) and/or longer trains of data are 
required.

We are very grateful to the staff members of the Goddard and VILSPA
IUE observatories for their assistance in scheduling and executing
these demanding monitoring programs. We also wish to thank our many
colleagues, including those on the IUE peer review committees, for 
their support of these programs. We gratefully acknowledge financial 
support of this particular program through an ADP grant
(NASA P.O. S-30917-F) to Computer Sciences Corporation.

\topmargin=-0.5in
\textheight=8.5in
\newpage
\scriptsize
\begin{center}
\begin{tabular}{lrrrrrrrrr}
\multicolumn{10}{c}{TABLE 1}
\\[0.2cm]
\multicolumn{10}{c}{GAUSSIAN COMPONENTS -- RANGES, MEANS, AND STANDARD
DEVIATIONS}
\\[0.2cm]
\hline
\hline
\multicolumn{1}{c}{Fitted}&
\multicolumn{3}{c}{F$_{0}$ (10$^{-14}$ ergs s$^{-1}$ cm$^{-2}$ \AA$^{-1}$)} &
\multicolumn{3}{c}{$\lambda_{c}$ (\AA) } &
\multicolumn{3}{c}{$\sigma_G$ (\AA) }  \\
\multicolumn{1}{c}{Component}&
\multicolumn{1}{c}{Min}& \multicolumn{1}{c}{Mean}& \multicolumn{1}{c}{Max}&
\multicolumn{1}{c}{Min}& \multicolumn{1}{c}{Mean}& \multicolumn{1}{c}{Max}&
\multicolumn{1}{c}{Min}& \multicolumn{1}{c}{Mean}& \multicolumn{1}{c}{Max}
\\ \hline

Ly$\alpha$ - Absorption&
 -212.9& -112.0$\pm$   30.6&  -51.5&
 1211.0& 1215.5$\pm$    1.9& 1220.0&
    1.8&    4.4$\pm$    1.3&    8.0\\
Ly$\alpha$ - Narrow&
   95.2&  164.0$\pm$   29.5&  270.0&
 1212.8& 1217.5$\pm$    1.3& 1220.2&
    2.1&    3.4$\pm$    0.4&    4.5\\
Ly$\alpha$ - Broad&
   59.6&   87.6$\pm$   19.0&  172.5&
 1216.1& 1219.9$\pm$    1.3& 1222.0&
    8.7&   12.6$\pm$    1.5&   155\\
N V &
  -60.0&  -44.4$\pm$    8.8&  -20.1&
 1236.0& 1238.1$\pm$    1.0& 1240.0&
    2.6&    3.9$\pm$    0.4&    5.0\\ \hline
C II - Absorption&
  -14.7&   -9.3$\pm$    2.1&   -5.5&
 1327.0& 1330.7$\pm$    1.3& 1333.6&
    1.5&    2.9$\pm$    0.7&    5.0\\
Si IV - Absorption 1&
  -32.5&  -23.5$\pm$    3.8&  -14.2&
 1387.0& 1389.5$\pm$    0.9& 1392.0&
    1.7&    3.1$\pm$    0.6&    4.1\\
Si IV - Absorption 2 &
  -23.9&  -15.7$\pm$    3.0&   -5.8&
 1396.0& 1398.4$\pm$    0.8& 1401.0&
    1.3&    2.1$\pm$    0.3&    3.0\\
Si IV + O IV] - Emission&
    6.0&   12.6$\pm$    3.9&   17.5&
 1388.0& 1390.3$\pm$    1.8& 1395.5&
    6.0&   12.0$\pm$    2.6&   16.0\\ \hline
N IV] &
    2.0&    6.2$\pm$    1.9&   11.5&
 1480.0& 1486.1$\pm$    3.1& 1491.0&
    0.7&    6.6$\pm$    2.5&    9.0\\
C IV - Absorption&
 -110.0&  -89.6$\pm$    8.8&  -70.0&
 1543.0& 1544.9$\pm$    0.9& 1547.0&
    2.7&    3.3$\pm$    0.2&    4.0\\
C IV - Narrow&
   62.8&   78.2$\pm$    7.7&   99.6&
 1544.0& 1545.8$\pm$    0.8& 1547.7&
    7.5&    9.7$\pm$    1.1&   12.0\\
C IV - Broad&
   23.8&   38.7$\pm$    3.8&   50.5&
 1543.0& 1546.1$\pm$    1.6& 1551.0&
   27.5&   34.5$\pm$    2.1&   36.5\\
He II &
   16.0&   19.8$\pm$    1.9&   25.7&
 1636.0& 1638.1$\pm$    1.0& 1640.0&
    4.5&    8.4$\pm$    1.7&   10.5\\
O III]&
    7.0&   10.8$\pm$    1.5&   14.5&
 1657.4& 1662.8$\pm$    2.0& 1666.0&
    4.5&    8.9$\pm$    1.0&    9.5\\ \hline
Al III Absorption&
   -5.0&   -2.6$\pm$    1.0&    0.0&
 1848.0& 1854.1$\pm$    3.3& 1860.0&
    0.0&    4.6$\pm$    1.9&    8.0\\
Si III] &
    4.0&    7.3$\pm$    1.7&   10.5&
 1879.0& 1884.2$\pm$    2.8& 1890.0&
    3.9&    9.9$\pm$    1.2&   10.5\\
C III] - Narrow&
   13.9&   21.0$\pm$    3.0&   28.3&
 1901.0& 1903.4$\pm$    1.0& 1905.8&
    3.3&    4.9$\pm$    0.5&    6.1\\
C III] - Broad &
    3.0&    6.6$\pm$    2.8&   12.0&
 1903.0& 1907.3$\pm$    4.7& 1919.0&
    7.0&   13.3$\pm$    3.0&   17.0\\
C III] - RED&
    1.0&    3.8$\pm$    1.5&    8.0&
 1912.0& 1915.0$\pm$    1.1& 1918.0&
    0.7&    2.5$\pm$    0.9&    4.7\\ \hline
\end{tabular}
\end{center}

\newpage
\begin{center}
\scriptsize
\begin{tabular}{cccccccc}
\multicolumn{8}{c}{TABLE 2} 
\\[0.2cm]
\multicolumn{8}{c}{SWP FLUXES$^a$}
\\[0.2cm]
\hline
\hline
\\[0.01cm]
\multicolumn{1}{c}{Image}&  
\multicolumn{1}{c}{Julian Date}&  
\multicolumn{1}{c}{$F_{\lambda}(1275$\,\AA)}&
\multicolumn{1}{c}{$F_{\lambda}(1440$\,\AA)}&
\multicolumn{1}{c}{$F_{\lambda}(1820$\,\AA)}&
\multicolumn{1}{c}{F(C IV)}&
\multicolumn{1}{c}{F(He II)}&
\multicolumn{1}{c}{F(C III])}\\
\multicolumn{1}{c}{} &{(2,440,000+)} &&&&&& \\
\\[0.01cm]
\hline
SWP 49333 &9318.83535 &36.94$\pm$0.34 &33.77$\pm$0.51 &25.67$\pm$0.34 &38.49$\pm$0.68 &4.48$\pm$0.18 &5.38$\pm$0.11 \\  
SWP 49334 &9318.88138 &35.88$\pm$0.37 &32.69$\pm$0.54 &25.74$\pm$0.36 &39.19$\pm$0.60 &4.90$\pm$0.22 &5.29$\pm$0.13 \\  
SWP 49335 &9318.92616 &36.63$\pm$0.36 &32.25$\pm$0.52 &25.01$\pm$0.35 &39.62$\pm$0.58 &5.00$\pm$0.24 &5.30$\pm$0.15 \\  
SWP 49341 &9319.81420 &36.27$\pm$0.35 &33.17$\pm$0.52 &26.05$\pm$0.36 &39.83$\pm$0.54 &4.68$\pm$0.33 &5.67$\pm$0.18 \\  
SWP 49342 &9319.85965 &38.64$\pm$0.37 &34.15$\pm$0.53 &26.84$\pm$0.36 &39.37$\pm$0.63 &4.50$\pm$0.37 &5.72$\pm$0.24 \\  
SWP 49343 &9319.90690 &38.45$\pm$0.36 &34.43$\pm$0.51 &26.42$\pm$0.33 &39.26$\pm$0.61 &4.70$\pm$0.35 &5.58$\pm$0.25 \\  
SWP 49362 &9320.84159 &37.98$\pm$0.37 &34.53$\pm$0.53 &27.91$\pm$0.35 &38.76$\pm$0.55 &5.29$\pm$0.23 &5.57$\pm$0.20 \\  
SWP 49363 &9320.88766 &38.74$\pm$0.37 &36.22$\pm$0.54 &27.30$\pm$0.35 &39.16$\pm$0.54 &5.84$\pm$0.19 &5.81$\pm$0.22 \\  
SWP 49364 &9320.93306 &38.50$\pm$0.36 &35.37$\pm$0.53 &27.29$\pm$0.36 &39.49$\pm$0.68 &5.69$\pm$0.18 &5.95$\pm$0.20 \\  
SWP 49372 &9321.83269 &37.33$\pm$0.36 &33.52$\pm$0.52 &27.03$\pm$0.36 &39.92$\pm$0.73 &5.37$\pm$0.19 &5.70$\pm$0.22 \\  
SWP 49373 &9321.88064 &38.89$\pm$0.36 &34.12$\pm$0.51 &27.00$\pm$0.34 &40.72$\pm$0.73 &5.24$\pm$0.19 &5.54$\pm$0.19 \\  
SWP 49374 &9321.92558 &38.96$\pm$0.36 &35.14$\pm$0.52 &28.04$\pm$0.36 &41.32$\pm$0.84 &5.50$\pm$0.20 &5.50$\pm$0.25 \\  
SWP 49379 &9322.65037 &40.47$\pm$0.37 &36.60$\pm$0.54 &27.52$\pm$0.35 &42.46$\pm$0.89 &5.53$\pm$0.21 &5.58$\pm$0.28 \\  
SWP 49380 &9322.70188 &40.82$\pm$0.38 &35.48$\pm$0.54 &27.12$\pm$0.35 &42.60$\pm$0.84 &5.61$\pm$0.21 &5.62$\pm$0.31 \\  
SWP 49381 &9322.74602 &39.94$\pm$0.38 &36.52$\pm$0.54 &28.53$\pm$0.37 &42.00$\pm$0.65 &5.89$\pm$0.21 &5.78$\pm$0.23 \\  
SWP 49382 &9322.79447 &40.78$\pm$0.38 &37.10$\pm$0.54 &29.16$\pm$0.36 &42.41$\pm$0.57 &6.15$\pm$0.21 &5.82$\pm$0.28 \\  
SWP 49383 &9322.83888 &43.33$\pm$0.39 &37.15$\pm$0.54 &28.81$\pm$0.35 &42.79$\pm$0.62 &6.26$\pm$0.24 &5.56$\pm$0.21 \\  
SWP 49384 &9322.91469 &44.40$\pm$0.40 &37.95$\pm$0.56 &29.30$\pm$0.37 &43.16$\pm$0.49 &6.06$\pm$0.35 &5.39$\pm$0.21 \\  
SWP 49385 &9322.96032 &44.46$\pm$0.40 &38.17$\pm$0.55 &28.89$\pm$0.37 &42.97$\pm$0.60 &6.15$\pm$0.58 &5.51$\pm$0.21 \\  
SWP 49386 &9323.01425 &44.62$\pm$0.40 &37.26$\pm$0.54 &28.65$\pm$0.37 &42.79$\pm$0.53 &6.21$\pm$0.57 &5.88$\pm$0.36 \\  
SWP 49387 &9323.06480 &44.77$\pm$0.40 &38.53$\pm$0.55 &29.28$\pm$0.37 &43.26$\pm$0.60 &6.30$\pm$0.47 &6.01$\pm$0.40 \\  
SWP 49388 &9323.11355 &44.19$\pm$0.40 &37.95$\pm$0.54 &28.96$\pm$0.37 &42.59$\pm$0.63 &6.20$\pm$0.22 &5.70$\pm$0.32 \\  
SWP 49389 &9323.16836 &44.75$\pm$0.40 &38.15$\pm$0.55 &30.17$\pm$0.36 &42.11$\pm$0.74 &6.06$\pm$0.21 &5.92$\pm$0.20 \\  
SWP 49390 &9323.24789 &42.06$\pm$0.38 &38.80$\pm$0.55 &28.29$\pm$0.36 &41.41$\pm$0.68 &5.95$\pm$0.22 &5.95$\pm$0.20 \\  
SWP 49391 &9323.29455 &42.95$\pm$0.38 &39.81$\pm$0.55 &29.06$\pm$0.35 &41.97$\pm$0.69 &5.93$\pm$0.31 &6.02$\pm$0.20 \\  
SWP 49392 &9323.34425 &44.26$\pm$0.39 &37.46$\pm$0.54 &29.02$\pm$0.35 &42.25$\pm$0.86 &6.18$\pm$0.60 &5.62$\pm$0.17 \\  
SWP 49393 &9323.38736 &43.13$\pm$0.39 &37.70$\pm$0.55 &29.19$\pm$0.36 &43.03$\pm$1.02 &6.37$\pm$0.58 &5.68$\pm$0.14 \\  
SWP 49395 &9323.57773 &42.90$\pm$0.39 &37.13$\pm$0.55 &28.82$\pm$0.35 &43.06$\pm$0.90 &6.48$\pm$0.54 &5.62$\pm$0.10 \\  
SWP 49396 &9323.62746 &42.76$\pm$0.39 &36.19$\pm$0.53 &29.21$\pm$0.35 &43.59$\pm$0.72 &6.21$\pm$0.23 &5.72$\pm$0.09 \\  
SWP 49397 &9323.67795 &45.04$\pm$0.39 &37.76$\pm$0.55 &29.03$\pm$0.36 &43.87$\pm$0.60 &5.71$\pm$0.20 &5.78$\pm$0.10 \\  
SWP 49398 &9323.72733 &43.03$\pm$0.39 &37.63$\pm$0.55 &29.49$\pm$0.37 &43.80$\pm$0.64 &5.46$\pm$0.20 &5.78$\pm$0.12 \\  
SWP 49399 &9323.77715 &44.87$\pm$0.40 &38.05$\pm$0.54 &30.72$\pm$0.36 &44.08$\pm$0.59 &5.73$\pm$0.20 &5.96$\pm$0.16 \\  
SWP 49400 &9323.82148 &44.56$\pm$0.39 &38.83$\pm$0.55 &29.58$\pm$0.36 &44.74$\pm$0.62 &6.02$\pm$0.22 &5.67$\pm$0.22 \\  
SWP 49401 &9323.89205 &43.17$\pm$0.38 &36.52$\pm$0.54 &29.14$\pm$0.35 &44.94$\pm$0.84 &6.40$\pm$0.20 &5.56$\pm$0.21 \\  
SWP 49402 &9323.93686 &42.46$\pm$0.39 &37.87$\pm$0.55 &29.62$\pm$0.36 &44.38$\pm$0.83 &6.56$\pm$0.19 &5.48$\pm$0.19 \\  
SWP 49403 &9323.98333 &42.72$\pm$0.39 &38.53$\pm$0.56 &28.99$\pm$0.37 &43.36$\pm$0.86 &6.28$\pm$0.19 &5.82$\pm$0.14 \\  
SWP 49404 &9324.03740 &42.48$\pm$0.39 &37.01$\pm$0.54 &28.95$\pm$0.37 &43.56$\pm$0.60 &6.34$\pm$0.26 &6.27$\pm$0.15 \\  
SWP 49405 &9324.08808 &43.58$\pm$0.40 &37.51$\pm$0.54 &29.28$\pm$0.37 &44.11$\pm$0.66 &6.06$\pm$0.28 &6.27$\pm$0.17 \\  
SWP 49406 &9324.14307 &44.02$\pm$0.40 &36.08$\pm$0.54 &27.86$\pm$0.37 &43.98$\pm$0.64 &6.58$\pm$0.28 &5.93$\pm$0.15 \\  
SWP 49407 &9324.19147 &42.49$\pm$0.40 &36.71$\pm$0.55 &28.64$\pm$0.38 &43.62$\pm$0.64 &6.49$\pm$0.22 &5.88$\pm$0.14 \\  
SWP 49408 &9324.25170 &43.40$\pm$0.39 &37.95$\pm$0.54 &28.67$\pm$0.35 &42.58$\pm$0.53 &6.83$\pm$0.18 &5.62$\pm$0.14 \\  
SWP 49409 &9324.29414 &43.87$\pm$0.40 &37.74$\pm$0.54 &28.98$\pm$0.36 &43.00$\pm$0.53 &6.71$\pm$0.19 &5.86$\pm$0.15 \\  
SWP 49410 &9324.33682 &42.18$\pm$0.39 &37.40$\pm$0.54 &27.94$\pm$0.37 &43.25$\pm$0.48 &6.56$\pm$0.19 &6.01$\pm$0.24 \\  
SWP 49411 &9324.38025 &40.80$\pm$0.38 &38.20$\pm$0.55 &28.60$\pm$0.37 &43.68$\pm$0.57 &6.45$\pm$0.24 &6.18$\pm$0.20 \\  
SWP 49412 &9324.42285 &40.96$\pm$0.38 &36.84$\pm$0.54 &28.82$\pm$0.36 &43.60$\pm$0.68 &6.10$\pm$0.30 &6.02$\pm$0.20 \\  
SWP 49413 &9324.56161 &40.54$\pm$0.38 &37.55$\pm$0.53 &28.47$\pm$0.35 &43.50$\pm$0.65 &5.77$\pm$0.29 &5.88$\pm$0.13 \\  
SWP 49414 &9324.60663 &40.99$\pm$0.38 &36.94$\pm$0.53 &28.32$\pm$0.36 &42.83$\pm$0.78 &5.86$\pm$0.28 &5.88$\pm$0.14 \\  
SWP 49415 &9324.65139 &42.83$\pm$0.39 &38.18$\pm$0.55 &27.92$\pm$0.36 &43.48$\pm$0.74 &6.15$\pm$0.24 &5.93$\pm$0.13 \\  
SWP 49416 &9324.69515 &40.93$\pm$0.39 &36.25$\pm$0.56 &27.91$\pm$0.37 &43.86$\pm$0.75 &6.56$\pm$0.27 &5.77$\pm$0.18 \\  
SWP 49417 &9324.73948 &40.84$\pm$0.39 &36.54$\pm$0.55 &28.00$\pm$0.38 &44.93$\pm$0.60 &6.50$\pm$0.24 &5.82$\pm$0.28 \\  
SWP 49418 &9324.78458 &40.18$\pm$0.38 &35.57$\pm$0.53 &28.06$\pm$0.35 &43.57$\pm$0.56 &6.38$\pm$0.23 &5.91$\pm$0.29 \\  
SWP 49419 &9324.83903 &40.05$\pm$0.37 &37.25$\pm$0.54 &28.23$\pm$0.35 &43.92$\pm$0.60 &6.34$\pm$0.33 &6.14$\pm$0.27 \\  
SWP 49420 &9324.89620 &38.89$\pm$0.37 &35.22$\pm$0.53 &28.41$\pm$0.35 &44.32$\pm$0.68 &6.20$\pm$0.34 &6.37$\pm$0.21 \\  
SWP 49421 &9324.94267 &40.48$\pm$0.38 &34.97$\pm$0.53 &27.98$\pm$0.35 &44.72$\pm$0.69 &6.51$\pm$0.36 &6.77$\pm$0.20 \\  
SWP 49422 &9324.98905 &39.73$\pm$0.38 &35.68$\pm$0.53 &27.22$\pm$0.35 &44.38$\pm$0.71 &6.11$\pm$0.25 &6.53$\pm$0.16 \\  
SWP 49423 &9325.04013 &38.47$\pm$0.36 &34.92$\pm$0.53 &27.95$\pm$0.37 &44.06$\pm$0.65 &5.69$\pm$0.25 &6.05$\pm$0.21 \\  
SWP 49424 &9325.08877 &38.85$\pm$0.37 &35.57$\pm$0.54 &27.35$\pm$0.35 &44.30$\pm$0.67 &5.98$\pm$0.24 &5.37$\pm$0.21 \\  
SWP 49425 &9325.13619 &39.25$\pm$0.37 &35.18$\pm$0.53 &27.93$\pm$0.36 &44.11$\pm$0.65 &6.47$\pm$0.24 &6.01$\pm$0.26 \\  

\end{tabular}

\begin{tabular}{cccccccc}
\multicolumn{8}{c}{TABLE 2 -- {\it Continued}}\\
\\[0.2cm]
\hline
\hline
\\[0.01cm]
\multicolumn{1}{c}{Image}&    
\multicolumn{1}{c}{Julian Date}&  
\multicolumn{1}{c}{$F_{\lambda}(1275$\,\AA)}&
\multicolumn{1}{c}{$F_{\lambda}(1440$\,\AA)}&
\multicolumn{1}{c}{$F_{\lambda}(1820$\,\AA)}&
\multicolumn{1}{c}{F(C IV)}&
\multicolumn{1}{c}{F(He II)}&
\multicolumn{1}{c}{F(C III])}\\
\multicolumn{1}{c}{} &{(2,440,000+)} &&&&&& \\
\\[0.01cm]
\hline

SWP 49426 &9325.18171 &40.78$\pm$0.39 &35.55$\pm$0.54 &27.22$\pm$0.36 &43.72$\pm$0.64 &6.53$\pm$0.23 &6.26$\pm$0.22 \\ 
SWP 49427 &9325.24969 &38.40$\pm$0.38 &36.69$\pm$0.55 &28.28$\pm$0.37 &44.11$\pm$0.52 &6.28$\pm$0.22 &6.77$\pm$0.23 \\ 
SWP 49429 &9325.35080 &39.54$\pm$0.36 &36.11$\pm$0.51 &28.09$\pm$0.35 &44.02$\pm$0.43 &5.99$\pm$0.21 &6.29$\pm$0.22 \\ 
SWP 49430 &9325.40051 &40.56$\pm$0.37 &37.03$\pm$0.54 &28.54$\pm$0.37 &44.78$\pm$0.38 &6.10$\pm$0.26 &6.30$\pm$0.21 \\ 
SWP 49431 &9325.54334 &40.26$\pm$0.38 &38.19$\pm$0.54 &29.18$\pm$0.36 &44.76$\pm$0.45 &6.10$\pm$0.27 &6.23$\pm$0.21 \\ 
SWP 49432 &9325.58506 &43.89$\pm$0.39 &39.03$\pm$0.55 &29.38$\pm$0.37 &45.18$\pm$0.56 &6.59$\pm$0.33 &6.40$\pm$0.26 \\ 
SWP 49433 &9325.62936 &44.22$\pm$0.40 &39.37$\pm$0.56 &29.18$\pm$0.37 &46.05$\pm$0.69 &7.09$\pm$0.28 &6.42$\pm$0.25 \\ 
SWP 49434 &9325.67491 &44.49$\pm$0.47 &39.89$\pm$0.69 &29.14$\pm$0.47 &45.77$\pm$0.79 &7.35$\pm$0.29 &6.34$\pm$0.33 \\ 
SWP 49435 &9325.71769 &46.26$\pm$0.51 &39.38$\pm$0.72 &30.02$\pm$0.51 &45.84$\pm$0.81 &7.22$\pm$0.24 &6.24$\pm$0.26 \\ 
SWP 49436 &9325.76265 &46.18$\pm$0.44 &39.56$\pm$0.60 &29.99$\pm$0.42 &44.19$\pm$0.78 &7.36$\pm$0.24 &6.28$\pm$0.32 \\ 
SWP 49437 &9325.80630 &45.30$\pm$0.44 &40.89$\pm$0.61 &29.07$\pm$0.40 &44.93$\pm$0.79 &7.13$\pm$0.20 &6.21$\pm$0.25 \\ 
SWP 49438 &9325.87929 &43.74$\pm$0.42 &38.75$\pm$0.58 &30.03$\pm$0.38 &44.55$\pm$0.77 &7.22$\pm$0.19 &6.22$\pm$0.24 \\ 
SWP 49439 &9325.92424 &45.82$\pm$0.42 &38.96$\pm$0.58 &30.02$\pm$0.39 &44.85$\pm$0.77 &6.87$\pm$0.20 &6.20$\pm$0.17 \\ 
SWP 49440 &9325.96878 &46.67$\pm$0.42 &39.78$\pm$0.59 &30.29$\pm$0.39 &44.81$\pm$0.67 &6.80$\pm$0.26 &6.47$\pm$0.14 \\ 
SWP 49441 &9326.01227 &45.57$\pm$0.42 &39.34$\pm$0.59 &30.09$\pm$0.40 &45.72$\pm$0.57 &6.38$\pm$0.28 &6.85$\pm$0.22 \\ 
SWP 49442 &9326.05646 &47.47$\pm$0.43 &38.45$\pm$0.59 &29.17$\pm$0.40 &45.87$\pm$0.55 &6.28$\pm$0.31 &6.97$\pm$0.25 \\ 
SWP 49443 &9326.10977 &46.90$\pm$0.43 &40.38$\pm$0.60 &30.65$\pm$0.40 &45.21$\pm$0.57 &6.10$\pm$0.46 &7.00$\pm$0.31 \\ 
SWP 49444 &9326.15278 &48.47$\pm$0.45 &42.31$\pm$0.62 &30.24$\pm$0.40 &45.24$\pm$0.57 &6.57$\pm$0.47 &6.55$\pm$0.27 \\ 
SWP 49445 &9326.19659 &49.80$\pm$0.45 &41.71$\pm$0.61 &32.03$\pm$0.42 &46.45$\pm$0.57 &6.52$\pm$0.46 &6.66$\pm$0.27 \\ 
SWP 49446 &9326.25544 &50.09$\pm$0.44 &42.74$\pm$0.60 &32.27$\pm$0.41 &46.95$\pm$0.63 &6.72$\pm$0.24 &6.75$\pm$0.24 \\ 
SWP 49447 &9326.29883 &50.15$\pm$0.43 &43.76$\pm$0.59 &32.46$\pm$0.39 &46.25$\pm$0.74 &6.79$\pm$0.20 &7.11$\pm$0.20 \\ 
SWP 49448 &9326.34414 &49.78$\pm$0.44 &42.86$\pm$0.61 &31.66$\pm$0.41 &45.11$\pm$0.76 &7.19$\pm$0.19 &7.05$\pm$0.21 \\ 
SWP 49449 &9326.38745 &51.45$\pm$0.46 &43.07$\pm$0.63 &32.51$\pm$0.41 &45.78$\pm$0.70 &7.13$\pm$0.30 &7.12$\pm$0.20 \\ 
SWP 49450 &9326.54594 &51.54$\pm$0.45 &41.72$\pm$0.60 &32.27$\pm$0.40 &45.58$\pm$0.72 &7.21$\pm$0.29 &6.90$\pm$0.20 \\ 
SWP 49451 &9326.58498 &51.92$\pm$0.43 &43.35$\pm$0.57 &31.62$\pm$0.38 &46.94$\pm$0.78 &7.40$\pm$0.51 &7.04$\pm$0.28 \\ 
SWP 49452 &9326.62965 &51.47$\pm$0.48 &44.34$\pm$0.65 &32.71$\pm$0.45 &46.47$\pm$0.82 &7.55$\pm$0.47 &7.20$\pm$0.28 \\ 
SWP 49453 &9326.67189 &52.30$\pm$0.52 &44.99$\pm$0.73 &32.64$\pm$0.51 &47.71$\pm$0.89 &7.20$\pm$0.49 &7.88$\pm$0.35 \\ 
SWP 49454 &9326.71376 &53.92$\pm$0.58 &44.25$\pm$0.80 &32.78$\pm$0.57 &47.38$\pm$0.92 &7.00$\pm$0.33 &7.82$\pm$0.23 \\ 
SWP 49455 &9326.75653 &54.34$\pm$0.50 &44.79$\pm$0.68 &32.78$\pm$0.46 &47.79$\pm$0.85 &7.11$\pm$0.28 &7.71$\pm$0.24 \\ 
SWP 49456 &9326.80056 &52.24$\pm$0.44 &43.07$\pm$0.57 &31.38$\pm$0.39 &47.14$\pm$0.73 &7.27$\pm$0.41 &7.22$\pm$0.27 \\ 
SWP 49457 &9326.85309 &52.83$\pm$0.44 &43.80$\pm$0.57 &31.92$\pm$0.37 &46.76$\pm$0.67 &7.16$\pm$0.34 &7.15$\pm$0.27 \\ 
SWP 49458 &9326.92141 &52.52$\pm$0.46 &43.02$\pm$0.62 &30.89$\pm$0.39 &46.19$\pm$0.66 &7.03$\pm$0.35 &7.24$\pm$0.31 \\ 
SWP 49459 &9326.97450 &51.37$\pm$0.45 &44.24$\pm$0.62 &31.01$\pm$0.41 &46.87$\pm$0.70 &6.72$\pm$0.26 &7.16$\pm$0.22 \\ 
SWP 49460 &9327.02091 &51.17$\pm$0.44 &42.85$\pm$0.61 &31.63$\pm$0.39 &46.78$\pm$0.62 &6.64$\pm$0.20 &7.06$\pm$0.26 \\ 
SWP 49461 &9327.06616 &51.00$\pm$0.44 &43.26$\pm$0.61 &31.33$\pm$0.38 &47.69$\pm$0.58 &6.68$\pm$0.20 &6.90$\pm$0.20 \\ 
SWP 49462 &9327.11621 &51.17$\pm$0.45 &43.05$\pm$0.61 &30.88$\pm$0.41 &47.93$\pm$0.56 &7.01$\pm$0.21 &7.14$\pm$0.28 \\ 
SWP 49463 &9327.16354 &49.42$\pm$0.44 &41.89$\pm$0.61 &30.63$\pm$0.39 &47.49$\pm$0.61 &7.15$\pm$0.26 &7.32$\pm$0.28 \\ 
SWP 49464 &9327.22521 &51.86$\pm$0.46 &44.27$\pm$0.62 &31.99$\pm$0.42 &47.53$\pm$0.72 &7.38$\pm$0.27 &7.16$\pm$0.26 \\ 
SWP 49465 &9327.26920 &51.15$\pm$0.48 &40.48$\pm$0.63 &31.94$\pm$0.42 &47.08$\pm$0.79 &7.32$\pm$0.24 &6.86$\pm$0.24 \\ 
SWP 49466 &9327.32016 &49.79$\pm$0.46 &41.82$\pm$0.63 &31.77$\pm$0.42 &48.08$\pm$0.99 &7.61$\pm$0.22 &6.75$\pm$0.26 \\ 
SWP 49467 &9327.37336 &49.57$\pm$0.47 &40.09$\pm$0.63 &30.33$\pm$0.42 &47.97$\pm$0.85 &7.26$\pm$0.24 &6.79$\pm$0.34 \\ 
SWP 49468 &9327.41718 &48.81$\pm$0.44 &40.46$\pm$0.62 &30.55$\pm$0.40 &47.09$\pm$0.85 &7.23$\pm$0.32 &6.84$\pm$0.35 \\ 
SWP 49470 &9327.53474 &47.47$\pm$0.44 &39.41$\pm$0.62 &29.22$\pm$0.40 &46.57$\pm$0.67 &6.68$\pm$0.34 &6.94$\pm$0.32 \\ 
SWP 49471 &9327.57271 &47.10$\pm$0.44 &39.23$\pm$0.60 &29.87$\pm$0.40 &46.72$\pm$0.78 &6.70$\pm$0.31 &6.70$\pm$0.23 \\ 
SWP 49472 &9327.61983 &47.01$\pm$0.46 &39.15$\pm$0.63 &30.42$\pm$0.44 &46.96$\pm$0.89 &6.55$\pm$0.28 &6.74$\pm$0.21 \\ 
SWP 49473 &9327.66413 &47.94$\pm$0.50 &41.75$\pm$0.73 &29.56$\pm$0.51 &48.11$\pm$0.89 &7.04$\pm$0.29 &7.03$\pm$0.24 \\ 
SWP 49474 &9327.70602 &48.37$\pm$0.55 &40.86$\pm$0.79 &29.61$\pm$0.55 &48.80$\pm$0.87 &6.88$\pm$0.30 &7.15$\pm$0.27 \\ 
SWP 49475 &9327.75097 &47.19$\pm$0.47 &40.05$\pm$0.65 &30.05$\pm$0.46 &49.19$\pm$0.66 &7.29$\pm$0.26 &7.24$\pm$0.20 \\ 
SWP 49476 &9327.79381 &45.03$\pm$0.40 &38.57$\pm$0.55 &30.33$\pm$0.37 &48.26$\pm$0.52 &6.94$\pm$0.21 &6.95$\pm$0.18 \\ 
SWP 49477 &9327.84021 &43.71$\pm$0.39 &39.64$\pm$0.55 &29.18$\pm$0.36 &47.38$\pm$0.52 &7.25$\pm$0.21 &6.72$\pm$0.22 \\ 
SWP 49478 &9327.89870 &43.53$\pm$0.39 &36.42$\pm$0.54 &28.92$\pm$0.39 &48.02$\pm$0.54 &7.00$\pm$0.18 &6.70$\pm$0.20 \\ 
SWP 49479 &9327.94343 &44.60$\pm$0.40 &35.76$\pm$0.53 &29.03$\pm$0.37 &48.83$\pm$0.57 &6.80$\pm$0.18 &6.59$\pm$0.23 \\ 
SWP 49480 &9327.99028 &44.04$\pm$0.39 &36.03$\pm$0.53 &29.09$\pm$0.36 &49.18$\pm$0.55 &6.70$\pm$0.19 &6.94$\pm$0.24 \\ 
SWP 49481 &9328.03307 &43.47$\pm$0.39 &36.94$\pm$0.54 &28.37$\pm$0.36 &48.56$\pm$0.54 &6.64$\pm$0.22 &6.67$\pm$0.25 \\ 
SWP 49482 &9328.08648 &43.17$\pm$0.38 &37.14$\pm$0.55 &28.97$\pm$0.37 &48.82$\pm$0.57 &6.97$\pm$0.26 &6.63$\pm$0.26 \\ 
SWP 49483 &9328.12875 &43.58$\pm$0.39 &36.71$\pm$0.54 &28.49$\pm$0.36 &48.61$\pm$0.67 &7.10$\pm$0.28 &6.61$\pm$0.21 \\ 
SWP 49484 &9328.17417 &44.02$\pm$0.41 &37.57$\pm$0.56 &29.56$\pm$0.38 &48.55$\pm$0.59 &7.22$\pm$0.29 &6.50$\pm$0.25 \\ 
SWP 49485 &9328.22543 &44.13$\pm$0.40 &39.36$\pm$0.57 &29.17$\pm$0.37 &47.93$\pm$0.58 &6.84$\pm$0.27 &6.78$\pm$0.32 \\ 

\end{tabular}

\begin{tabular}{cccccccc}
\multicolumn{8}{c}{TABLE 2 -- {\it Continued}}\\
\\[0.2cm]
\hline
\hline
\\[0.01cm]
\multicolumn{1}{c}{Image}&    
\multicolumn{1}{c}{Julian Date}&
\multicolumn{1}{c}{$F_{\lambda}(1275$\,\AA)}&
\multicolumn{1}{c}{$F_{\lambda}(1440$\,\AA)}&
\multicolumn{1}{c}{$F_{\lambda}(1820$\,\AA)}&
\multicolumn{1}{c}{F(C IV)}&
\multicolumn{1}{c}{F(He II)}&
\multicolumn{1}{c}{F(C III])}\\
\multicolumn{1}{c}{} &{(2,440,000+)} &&&&&& \\
\\[0.01cm]
\hline

SWP 49486 &9328.27341 &45.21$\pm$0.41 &37.89$\pm$0.55 &28.36$\pm$0.37 &48.30$\pm$0.48 &6.98$\pm$0.22 &6.69$\pm$0.36 \\ 
SWP 49487 &9328.31764 &44.23$\pm$0.42 &36.73$\pm$0.58 &29.20$\pm$0.39 &48.96$\pm$0.57 &6.84$\pm$0.22 &6.79$\pm$0.30 \\ 
SWP 49488 &9328.36374 &43.37$\pm$0.41 &36.78$\pm$0.57 &28.65$\pm$0.39 &49.16$\pm$0.61 &7.12$\pm$0.21 &6.42$\pm$0.16 \\ 
SWP 49489 &9328.40705 &42.54$\pm$0.41 &36.43$\pm$0.57 &28.37$\pm$0.39 &47.95$\pm$0.59 &6.75$\pm$0.23 &6.23$\pm$0.12 \\ 
SWP 49491 &9328.54118 &41.44$\pm$0.39 &36.09$\pm$0.55 &28.39$\pm$0.36 &47.40$\pm$0.60 &6.57$\pm$0.43 &6.25$\pm$0.13 \\ 
SWP 49492 &9328.58108 &42.24$\pm$0.38 &37.51$\pm$0.53 &29.15$\pm$0.35 &46.86$\pm$0.73 &6.67$\pm$0.46 &6.19$\pm$0.13 \\ 
SWP 49493 &9328.62321 &44.05$\pm$0.40 &39.17$\pm$0.56 &28.60$\pm$0.37 &46.39$\pm$0.75 &6.21$\pm$0.45 &6.54$\pm$0.13 \\ 
SWP 49494 &9328.66607 &44.99$\pm$0.45 &40.74$\pm$0.64 &26.70$\pm$0.42 &46.52$\pm$0.90 &6.23$\pm$0.51 &6.79$\pm$0.15 \\ 
SWP 49495 &9328.70771 &44.55$\pm$0.49 &39.58$\pm$0.70 &28.54$\pm$0.47 &47.16$\pm$0.84 &5.72$\pm$0.49 &7.28$\pm$0.23 \\ 
SWP 49496 &9328.75026 &45.09$\pm$0.43 &39.34$\pm$0.61 &28.59$\pm$0.41 &47.94$\pm$0.81 &6.34$\pm$0.48 &7.19$\pm$0.30 \\ 
SWP 49497 &9328.79306 &44.29$\pm$0.38 &39.06$\pm$0.52 &28.89$\pm$0.35 &47.51$\pm$0.62 &6.65$\pm$0.22 &6.94$\pm$0.35 \\ 
SWP 49498 &9328.84015 &44.15$\pm$0.37 &38.77$\pm$0.52 &29.33$\pm$0.34 &47.01$\pm$0.54 &6.84$\pm$0.22 &6.72$\pm$0.28 \\ 
SWP 49499 &9328.89331 &45.56$\pm$0.40 &38.92$\pm$0.55 &29.35$\pm$0.36 &48.30$\pm$0.62 &6.69$\pm$0.43 &6.61$\pm$0.27 \\ 
SWP 49500 &9328.93673 &44.60$\pm$0.40 &38.40$\pm$0.56 &29.54$\pm$0.38 &48.85$\pm$0.67 &6.55$\pm$0.42 &6.62$\pm$0.21 \\ 
SWP 49501 &9328.98288 &46.95$\pm$0.41 &37.86$\pm$0.55 &29.53$\pm$0.37 &48.88$\pm$0.65 &7.03$\pm$0.41 &6.84$\pm$0.24 \\ 
SWP 49502 &9329.02826 &46.42$\pm$0.41 &39.13$\pm$0.55 &30.19$\pm$0.38 &48.53$\pm$0.53 &7.30$\pm$0.20 &7.00$\pm$0.20 \\ 
SWP 49503 &9329.07162 &46.76$\pm$0.41 &40.09$\pm$0.56 &29.47$\pm$0.38 &48.05$\pm$0.51 &7.28$\pm$0.19 &6.83$\pm$0.20 \\ 
SWP 49504 &9329.11318 &48.05$\pm$0.42 &40.57$\pm$0.57 &31.07$\pm$0.38 &48.17$\pm$0.62 &6.94$\pm$0.22 &6.43$\pm$0.18 \\ 
SWP 49505 &9329.15711 &48.60$\pm$0.42 &39.76$\pm$0.56 &30.79$\pm$0.37 &47.55$\pm$0.71 &6.99$\pm$0.22 &6.23$\pm$0.19 \\ 
SWP 49506 &9329.21402 &48.63$\pm$0.42 &39.20$\pm$0.55 &30.49$\pm$0.37 &48.04$\pm$0.69 &7.29$\pm$0.23 &6.53$\pm$0.17 \\ 
SWP 49507 &9329.25431 &46.39$\pm$0.41 &40.41$\pm$0.58 &29.67$\pm$0.37 &48.02$\pm$0.57 &7.57$\pm$0.20 &6.71$\pm$0.27 \\ 
SWP 49508 &9329.30167 &47.39$\pm$0.42 &39.23$\pm$0.57 &31.00$\pm$0.38 &48.53$\pm$0.64 &7.23$\pm$0.22 &6.86$\pm$0.28 \\ 
SWP 49509 &9329.34620 &48.32$\pm$0.43 &38.12$\pm$0.57 &30.42$\pm$0.39 &47.28$\pm$0.64 &6.96$\pm$0.22 &6.59$\pm$0.28 \\ 
SWP 49510 &9329.39206 &46.53$\pm$0.40 &38.39$\pm$0.54 &30.24$\pm$0.36 &46.92$\pm$0.68 &6.61$\pm$0.22 &6.66$\pm$0.18 \\ 
SWP 49511 &9329.43838 &45.96$\pm$0.38 &37.24$\pm$0.52 &29.37$\pm$0.38 &45.84$\pm$0.61 &7.08$\pm$0.19 &6.64$\pm$0.16 \\ 
SWP 49512 &9329.53191 &43.49$\pm$0.40 &37.42$\pm$0.54 &28.62$\pm$0.35 &47.12$\pm$0.65 &7.21$\pm$0.20 &6.83$\pm$0.19 \\ 
SWP 49513 &9329.57171 &42.53$\pm$0.40 &38.08$\pm$0.56 &28.90$\pm$0.38 &47.38$\pm$0.79 &7.42$\pm$0.25 &6.81$\pm$0.31 \\ 
SWP 49514 &9329.61359 &41.83$\pm$0.39 &36.20$\pm$0.55 &28.13$\pm$0.37 &47.73$\pm$0.90 &6.98$\pm$0.27 &7.15$\pm$0.31 \\ 
SWP 49515 &9329.65601 &42.03$\pm$0.38 &36.54$\pm$0.53 &27.52$\pm$0.35 &47.13$\pm$0.83 &6.56$\pm$0.27 &7.09$\pm$0.30 \\ 
SWP 49516 &9329.69817 &41.61$\pm$0.40 &37.08$\pm$0.57 &28.47$\pm$0.39 &46.92$\pm$0.70 &6.57$\pm$0.21 &6.98$\pm$0.22 \\ 
SWP 49517 &9329.74183 &41.36$\pm$0.39 &35.82$\pm$0.56 &28.15$\pm$0.36 &47.67$\pm$0.62 &6.66$\pm$0.22 &6.88$\pm$0.23 \\ 
SWP 49518 &9329.78385 &41.00$\pm$0.39 &35.42$\pm$0.55 &27.98$\pm$0.37 &48.27$\pm$0.63 &6.77$\pm$0.24 &6.98$\pm$0.32 \\ 
SWP 49519 &9329.83279 &39.59$\pm$0.37 &35.42$\pm$0.53 &26.35$\pm$0.35 &47.72$\pm$0.78 &6.66$\pm$0.29 &7.06$\pm$0.29 \\ 
SWP 49520 &9329.88559 &39.88$\pm$0.38 &35.86$\pm$0.53 &27.86$\pm$0.35 &46.54$\pm$0.74 &6.58$\pm$0.30 &6.75$\pm$0.28 \\ 
SWP 49521 &9329.92980 &39.54$\pm$0.37 &35.97$\pm$0.53 &27.71$\pm$0.36 &45.57$\pm$0.83 &6.73$\pm$0.29 &6.72$\pm$0.14 \\ 
SWP 49522 &9329.97655 &40.39$\pm$0.37 &35.03$\pm$0.53 &26.92$\pm$0.36 &46.29$\pm$0.71 &6.71$\pm$0.26 &6.65$\pm$0.25 \\ 
SWP 49523 &9330.02228 &41.22$\pm$0.38 &35.18$\pm$0.54 &27.46$\pm$0.35 &45.75$\pm$0.72 &6.66$\pm$0.24 &6.37$\pm$0.25 \\ 
SWP 49524 &9330.07459 &39.21$\pm$0.38 &35.99$\pm$0.53 &27.65$\pm$0.35 &46.72$\pm$0.62 &6.49$\pm$0.22 &6.19$\pm$0.26 \\ 
SWP 49525 &9330.11640 &40.12$\pm$0.37 &34.36$\pm$0.53 &27.99$\pm$0.37 &46.60$\pm$0.61 &6.30$\pm$0.20 &6.25$\pm$0.16 \\ 
SWP 49526 &9330.15976 &39.59$\pm$0.38 &35.34$\pm$0.53 &28.89$\pm$0.36 &47.27$\pm$0.61 &6.47$\pm$0.19 &6.64$\pm$0.17 \\ 
SWP 49527 &9330.21123 &40.55$\pm$0.39 &35.67$\pm$0.54 &27.06$\pm$0.35 &46.62$\pm$0.75 &6.33$\pm$0.27 &6.69$\pm$0.24 \\ 
SWP 49528 &9330.25175 &39.23$\pm$0.37 &37.48$\pm$0.55 &28.43$\pm$0.36 &46.49$\pm$0.66 &6.46$\pm$0.30 &6.56$\pm$0.22 \\ 
SWP 49529 &9330.29745 &40.00$\pm$0.37 &36.35$\pm$0.54 &27.69$\pm$0.36 &46.99$\pm$0.60 &6.27$\pm$0.29 &6.75$\pm$0.23 \\ 
SWP 49530 &9330.34584 &41.40$\pm$0.39 &36.12$\pm$0.53 &27.55$\pm$0.35 &47.12$\pm$0.50 &6.25$\pm$0.24 &6.92$\pm$0.14 \\ 
SWP 49531 &9330.38880 &40.99$\pm$0.39 &35.30$\pm$0.56 &27.70$\pm$0.38 &47.38$\pm$0.58 &6.12$\pm$0.20 &7.01$\pm$0.15 \\ 
SWP 49533 &9330.52881 &40.17$\pm$0.37 &36.27$\pm$0.54 &27.31$\pm$0.36 &47.68$\pm$0.52 &5.90$\pm$0.19 &6.77$\pm$0.12 \\ 
SWP 49534 &9330.56726 &40.60$\pm$0.39 &36.10$\pm$0.55 &27.63$\pm$0.36 &47.17$\pm$0.55 &6.15$\pm$0.27 &6.55$\pm$0.15 \\ 
SWP 49535 &9330.61240 &40.62$\pm$0.39 &36.28$\pm$0.54 &28.20$\pm$0.36 &46.99$\pm$0.48 &6.25$\pm$0.29 &6.59$\pm$0.15 \\ 
SWP 49536 &9330.65634 &41.61$\pm$0.40 &36.11$\pm$0.57 &27.25$\pm$0.39 &47.06$\pm$0.59 &6.42$\pm$0.34 &6.28$\pm$0.16 \\ 
SWP 49537 &9330.70314 &42.72$\pm$0.42 &35.75$\pm$0.57 &28.00$\pm$0.39 &48.06$\pm$0.47 &6.65$\pm$0.26 &6.45$\pm$0.13 \\ 
SWP 49538 &9330.74627 &41.34$\pm$0.40 &35.37$\pm$0.56 &28.47$\pm$0.38 &48.06$\pm$0.69 &6.71$\pm$0.30 &6.69$\pm$0.15 \\ 
SWP 49539 &9330.79020 &41.48$\pm$0.38 &36.60$\pm$0.54 &27.40$\pm$0.35 &47.59$\pm$0.60 &6.58$\pm$0.29 &7.14$\pm$0.27 \\ 
SWP 49540 &9330.84397 &43.04$\pm$0.40 &36.15$\pm$0.54 &27.59$\pm$0.36 &47.97$\pm$0.66 &6.42$\pm$0.30 &7.15$\pm$0.27 \\ 
SWP 49541 &9330.90933 &42.04$\pm$0.39 &36.29$\pm$0.54 &28.65$\pm$0.37 &46.96$\pm$0.60 &6.59$\pm$0.29 &6.72$\pm$0.24 \\ 
SWP 49542 &9330.96370 &43.17$\pm$0.39 &37.70$\pm$0.55 &28.87$\pm$0.35 &47.06$\pm$0.82 &6.52$\pm$0.29 &6.84$\pm$0.12 \\ 
SWP 49543 &9331.00844 &43.52$\pm$0.40 &36.20$\pm$0.54 &27.81$\pm$0.36 &47.00$\pm$0.83 &6.32$\pm$0.28 &6.88$\pm$0.16 \\ 
SWP 49544 &9331.05978 &43.31$\pm$0.39 &35.28$\pm$0.53 &28.02$\pm$0.37 &48.14$\pm$0.74 &6.50$\pm$0.26 &7.16$\pm$0.22 \\ 
SWP 49545 &9331.10807 &43.32$\pm$0.40 &36.43$\pm$0.54 &28.72$\pm$0.37 &47.65$\pm$0.53 &6.75$\pm$0.25 &6.77$\pm$0.25 \\ 

\end{tabular}

\begin{tabular}{cccccccc}
\multicolumn{8}{c}{TABLE 2 -- {\it Continued}}\\
\\[0.2cm]
\hline
\hline
\\[0.01cm]
\multicolumn{1}{c}{Image}&    
\multicolumn{1}{c}{Julian Date}&
\multicolumn{1}{c}{$F_{\lambda}(1275$\,\AA)}&
\multicolumn{1}{c}{$F_{\lambda}(1440$\,\AA)}&
\multicolumn{1}{c}{$F_{\lambda}(1820$\,\AA)}&
\multicolumn{1}{c}{F(C IV)}&
\multicolumn{1}{c}{F(He II)}&
\multicolumn{1}{c}{F(C III])}\\
\multicolumn{1}{c}{} &{(2,440,000+)} &&&&&& \\
\\[0.01cm]
\hline

SWP 49546 &9331.15458 &43.62$\pm$0.40 &36.61$\pm$0.54 &27.57$\pm$0.35 &45.67$\pm$0.57 &6.59$\pm$0.23 &6.66$\pm$0.30 \\ 
SWP 49547 &9331.21365 &43.33$\pm$0.39 &39.14$\pm$0.53 &28.89$\pm$0.36 &46.42$\pm$0.52 &6.17$\pm$0.20 &6.64$\pm$0.22 \\ 
SWP 49548 &9331.25499 &42.40$\pm$0.38 &36.51$\pm$0.53 &28.49$\pm$0.36 &47.47$\pm$0.70 &6.22$\pm$0.18 &7.05$\pm$0.23 \\ 
SWP 49549 &9331.30012 &42.43$\pm$0.39 &37.21$\pm$0.53 &28.07$\pm$0.36 &49.74$\pm$0.77 &6.62$\pm$0.29 &6.85$\pm$0.16 \\ 
SWP 49550 &9331.34586 &42.94$\pm$0.38 &35.37$\pm$0.52 &28.85$\pm$0.38 &49.38$\pm$0.85 &6.57$\pm$0.29 &6.80$\pm$0.18 \\ 
SWP 49551 &9331.38575 &42.06$\pm$0.38 &36.13$\pm$0.54 &27.58$\pm$0.34 &47.93$\pm$0.72 &6.91$\pm$0.28 &6.39$\pm$0.23 \\ 
SWP 49553 &9331.51858 &39.24$\pm$0.38 &35.27$\pm$0.54 &26.32$\pm$0.36 &47.15$\pm$0.53 &7.02$\pm$0.17 &6.45$\pm$0.22 \\ 
SWP 49554 &9331.55876 &38.16$\pm$0.36 &35.14$\pm$0.54 &27.21$\pm$0.36 &46.50$\pm$0.58 &7.32$\pm$0.17 &6.51$\pm$0.29 \\ 
SWP 49555 &9331.60369 &40.15$\pm$0.38 &35.14$\pm$0.53 &27.78$\pm$0.36 &47.06$\pm$0.66 &6.99$\pm$0.33 &6.92$\pm$0.28 \\ 
SWP 49556 &9331.64840 &39.85$\pm$0.40 &36.62$\pm$0.57 &27.02$\pm$0.37 &47.80$\pm$0.94 &7.10$\pm$0.36 &7.34$\pm$0.29 \\ 
SWP 49557 &9331.69266 &39.22$\pm$0.40 &35.56$\pm$0.59 &27.18$\pm$0.41 &47.84$\pm$0.99 &7.03$\pm$0.37 &7.59$\pm$0.24 \\ 
SWP 49558 &9331.73834 &37.72$\pm$0.39 &33.90$\pm$0.55 &26.24$\pm$0.38 &48.53$\pm$0.92 &6.91$\pm$0.25 &7.37$\pm$0.19 \\ 
SWP 49559 &9331.80425 &36.86$\pm$0.39 &32.47$\pm$0.55 &26.44$\pm$0.36 &48.13$\pm$0.71 &6.31$\pm$0.23 &7.05$\pm$0.18 \\ 
SWP 49560 &9331.85256 &37.09$\pm$0.37 &30.99$\pm$0.53 &25.94$\pm$0.36 &48.12$\pm$0.71 &6.19$\pm$0.27 &6.61$\pm$0.18 \\ 
SWP 49567 &9332.75464 &42.61$\pm$0.38 &36.73$\pm$0.55 &27.48$\pm$0.35 &48.88$\pm$0.67 &6.51$\pm$0.24 &6.57$\pm$0.25 \\ 
SWP 49568 &9332.80381 &41.76$\pm$0.38 &36.39$\pm$0.54 &29.26$\pm$0.36 &48.60$\pm$0.62 &7.07$\pm$0.23 &6.88$\pm$0.30 \\ 
SWP 49569 &9332.85104 &42.67$\pm$0.39 &36.40$\pm$0.54 &28.21$\pm$0.35 &48.66$\pm$0.48 &7.16$\pm$0.18 &6.74$\pm$0.31 \\ 
SWP 49574 &9333.74031 &38.47$\pm$0.36 &33.57$\pm$0.52 &27.45$\pm$0.37 &47.80$\pm$0.49 &7.07$\pm$0.21 &6.71$\pm$0.26 \\ 
SWP 49575 &9333.78330 &38.26$\pm$0.33 &34.03$\pm$0.47 &26.67$\pm$0.31 &48.21$\pm$0.61 &6.94$\pm$0.20 &6.55$\pm$0.24 \\ 
SWP 49576 &9333.82860 &38.13$\pm$0.32 &33.49$\pm$0.45 &26.55$\pm$0.30 &49.15$\pm$0.60 &6.85$\pm$0.20 &6.75$\pm$0.22 \\ 
SWP 49582 &9334.75142 &37.84$\pm$0.34 &32.84$\pm$0.46 &26.27$\pm$0.31 &49.13$\pm$0.58 &6.71$\pm$0.20 &6.66$\pm$0.18 \\ 
SWP 49583 &9334.79190 &37.45$\pm$0.35 &32.91$\pm$0.49 &26.46$\pm$0.32 &48.26$\pm$0.71 &6.47$\pm$0.50 &6.53$\pm$0.17 \\ 
SWP 49584 &9334.83455 &38.30$\pm$0.35 &35.54$\pm$0.49 &26.37$\pm$0.31 &47.20$\pm$0.73 &6.55$\pm$0.50 &6.81$\pm$0.17 \\ 
SWP 49592 &9335.74617 &42.78$\pm$0.36 &36.36$\pm$0.49 &27.49$\pm$0.33 &46.59$\pm$0.82 &6.52$\pm$0.52 &6.87$\pm$0.15 \\ 
SWP 49593 &9335.78609 &42.27$\pm$0.36 &36.11$\pm$0.49 &28.79$\pm$0.34 &47.50$\pm$0.66 &6.46$\pm$0.22 &7.00$\pm$0.15 \\ 
SWP 49594 &9335.83018 &41.66$\pm$0.36 &36.89$\pm$0.49 &27.81$\pm$0.33 &48.59$\pm$0.68 &6.60$\pm$0.22 &6.62$\pm$0.14 \\ 
SWP 49600 &9336.75053 &44.35$\pm$0.37 &37.02$\pm$0.50 &29.86$\pm$0.34 &49.99$\pm$0.58 &6.78$\pm$0.21 &6.51$\pm$0.15 \\ 
SWP 49601 &9336.79082 &42.41$\pm$0.35 &35.61$\pm$0.48 &28.08$\pm$0.32 &50.55$\pm$0.53 &7.16$\pm$0.19 &6.56$\pm$0.16 \\ 
SWP 49602 &9336.82708 &44.31$\pm$0.37 &36.54$\pm$0.50 &28.63$\pm$0.33 &51.06$\pm$0.48 &7.22$\pm$0.14 &7.11$\pm$0.22 \\ 
\\[0.01cm]
\hline
\multicolumn{7}{l}{
$^a$Rest-frame continuum fluxes in units of 10$^{-14}$ ergs
s$^{-1}$ cm$^{-2}$ \AA$^{-1}$.}\\
\multicolumn{7}{l}{
Rest-frame line fluxes in units of
10$^{-12}$ ergs s$^{-1}$ cm$^{-2}$.}\\
\end{tabular}
\end{center}

\newpage
\begin{center}
\scriptsize
\begin{tabular}{ccccccc}
\multicolumn{7}{c}{TABLE 3} 
\\[0.2cm]
\multicolumn{7}{c}{LWP FLUXES$^a$}
\\[0.2cm]
\hline
\hline
\\[0.01cm]
\multicolumn{1}{c}{Image} &
\multicolumn{1}{c}{Julian Date} &
\multicolumn{1}{c}{$F_{\lambda}(2688$\,\AA)} &{~~~} &
\multicolumn{1}{c}{Image} &
\multicolumn{1}{c}{Julian Date} &
\multicolumn{1}{c}{$F_{\lambda}(2688$\,\AA)} \\
\multicolumn{1}{c}{} & {(2,440,000+)} & & &
\multicolumn{1}{c}{} & {(2,440,000+)} & \\
\\[0.01cm]
\hline

LWP 26815 &9318.85602 &17.92$\pm$0.20 & &LWP 26893 &9325.23022 &19.15$\pm$0.20 \\ 
LWP 26816 &9318.90362 &17.75$\pm$0.18 & &LWP 26896 &9325.37193 &19.27$\pm$0.22 \\ 
LWP 26817 &9318.94549 &17.92$\pm$0.18 & &LWP 26897 &9325.42154 &19.44$\pm$0.22 \\ 
LWP 26822 &9319.83692 &17.83$\pm$0.18 & &LWP 26898 &9325.55997 &19.00$\pm$0.22 \\ 
LWP 26823 &9319.87985 &17.82$\pm$0.19 & &LWP 26899 &9325.60505 &19.20$\pm$0.22 \\ 
LWP 26824 &9319.92803 &17.41$\pm$0.19 & &LWP 26900 &9325.65110 &19.74$\pm$0.24 \\ 
LWP 26831 &9320.86331 &18.00$\pm$0.19 & &LWP 26901 &9325.69425 &20.04$\pm$0.28 \\ 
LWP 26832 &9320.91007 &18.12$\pm$0.17 & &LWP 26902 &9325.73752 &20.67$\pm$0.27 \\ 
LWP 26836 &9321.85620 &18.00$\pm$0.18 & &LWP 26903 &9325.78252 &19.54$\pm$0.23 \\ 
LWP 26837 &9321.90200 &18.39$\pm$0.18 & &LWP 26904 &9325.83486 &18.98$\pm$0.23 \\ 
LWP 26838 &9321.94529 &18.17$\pm$0.19 & &LWP 26905 &9325.89709 &19.79$\pm$0.23 \\ 
LWP 26846 &9322.67571 &18.48$\pm$0.20 & &LWP 26906 &9325.94435 &19.88$\pm$0.23 \\ 
LWP 26847 &9322.72296 &18.77$\pm$0.20 & &LWP 26907 &9325.98786 &19.92$\pm$0.23 \\ 
LWP 26848 &9322.76984 &18.76$\pm$0.20 & &LWP 26908 &9326.03159 &20.25$\pm$0.23 \\ 
LWP 26849 &9322.81449 &18.84$\pm$0.20 & &LWP 26909 &9326.08428 &19.72$\pm$0.24 \\ 
LWP 26850 &9322.88833 &19.36$\pm$0.20 & &LWP 26910 &9326.12891 &19.86$\pm$0.24 \\ 
LWP 26851 &9322.93730 &19.15$\pm$0.20 & &LWP 26911 &9326.17253 &20.16$\pm$0.24 \\ 
LWP 26852 &9322.98483 &19.17$\pm$0.20 & &LWP 26912 &9326.23507 &20.59$\pm$0.23 \\ 
LWP 26853 &9323.03668 &19.29$\pm$0.20 & &LWP 26913 &9326.27462 &20.25$\pm$0.23 \\ 
LWP 26854 &9323.08673 &19.09$\pm$0.20 & &LWP 26914 &9326.32039 &20.51$\pm$0.24 \\ 
LWP 26855 &9323.13972 &19.27$\pm$0.20 & &LWP 26915 &9326.36429 &20.43$\pm$0.25 \\ 
LWP 26856 &9323.22432 &19.86$\pm$0.20 & &LWP 26916 &9326.40734 &20.76$\pm$0.24 \\ 
LWP 26857 &9323.26862 &19.31$\pm$0.22 & &LWP 26917 &9326.56074 &20.82$\pm$0.24 \\ 
LWP 26858 &9323.31392 &19.39$\pm$0.22 & &LWP 26918 &9326.60631 &20.66$\pm$0.25 \\ 
LWP 26859 &9323.36330 &19.46$\pm$0.22 & &LWP 26919 &9326.64889 &20.82$\pm$0.26 \\ 
LWP 26860 &9323.40693 &19.40$\pm$0.22 & &LWP 26920 &9326.69127 &21.29$\pm$0.30 \\ 
LWP 26861 &9323.55218 &19.16$\pm$0.22 & &LWP 26921 &9326.73370 &20.86$\pm$0.30 \\ 
LWP 26862 &9323.59907 &19.22$\pm$0.22 & &LWP 26922 &9326.77600 &21.01$\pm$0.25 \\ 
LWP 26863 &9323.64899 &19.38$\pm$0.22 & &LWP 26923 &9326.81910 &20.64$\pm$0.24 \\ 
LWP 26864 &9323.69892 &19.49$\pm$0.22 & &LWP 26924 &9326.89598 &21.07$\pm$0.24 \\ 
LWP 26865 &9323.74873 &19.41$\pm$0.22 & &LWP 26925 &9326.94789 &20.74$\pm$0.23 \\ 
LWP 26866 &9323.79730 &19.60$\pm$0.22 & &LWP 26926 &9326.99395 &20.83$\pm$0.24 \\ 
LWP 26867 &9323.84314 &19.78$\pm$0.22 & &LWP 26927 &9327.04029 &20.95$\pm$0.24 \\ 
LWP 26868 &9323.91324 &19.18$\pm$0.20 & &LWP 26928 &9327.09056 &21.08$\pm$0.24 \\ 
LWP 26869 &9323.95897 &19.28$\pm$0.20 & &LWP 26929 &9327.13751 &20.84$\pm$0.24 \\ 
LWP 26870 &9324.00767 &19.42$\pm$0.20 & &LWP 26930 &9327.19061 &21.04$\pm$0.25 \\ 
LWP 26871 &9324.05895 &19.18$\pm$0.20 & &LWP 26932 &9327.29420 &20.97$\pm$0.24 \\ 
LWP 26872 &9324.11755 &18.95$\pm$0.20 & &LWP 26933 &9327.33920 &21.20$\pm$0.24 \\ 
LWP 26873 &9324.16673 &19.18$\pm$0.20 & &LWP 26934 &9327.39288 &20.78$\pm$0.25 \\ 
LWP 26874 &9324.23083 &19.14$\pm$0.22 & &LWP 26935 &9327.54927 &20.56$\pm$0.24 \\ 
LWP 26875 &9324.27104 &19.35$\pm$0.22 & &LWP 26936 &9327.59721 &20.56$\pm$0.24 \\ 
LWP 26876 &9324.31343 &19.40$\pm$0.22 & &LWP 26937 &9327.63990 &20.95$\pm$0.26 \\ 
LWP 26877 &9324.35664 &19.45$\pm$0.22 & &LWP 26938 &9327.68222 &21.57$\pm$0.33 \\ 
LWP 26878 &9324.39969 &19.31$\pm$0.22 & &LWP 26939 &9327.72759 &21.12$\pm$0.29 \\ 
LWP 26879 &9324.54413 &18.53$\pm$0.21 & &LWP 26940 &9327.77062 &20.38$\pm$0.23 \\ 
LWP 26880 &9324.58134 &18.88$\pm$0.21 & &LWP 26941 &9327.81279 &20.38$\pm$0.22 \\ 
LWP 26881 &9324.62566 &19.51$\pm$0.22 & &LWP 26942 &9327.87748 &20.40$\pm$0.22 \\ 
LWP 26882 &9324.67076 &18.79$\pm$0.22 & &LWP 26943 &9327.91932 &20.83$\pm$0.22 \\ 
LWP 26883 &9324.71550 &18.87$\pm$0.23 & &LWP 26944 &9327.96603 &20.37$\pm$0.22 \\ 
LWP 26884 &9324.76015 &18.99$\pm$0.22 & &LWP 26945 &9328.00942 &20.35$\pm$0.22 \\ 
LWP 26885 &9324.80494 &18.58$\pm$0.21 & &LWP 26946 &9328.06204 &20.54$\pm$0.22 \\ 
LWP 26886 &9324.87563 &19.21$\pm$0.20 & &LWP 26947 &9328.10593 &20.60$\pm$0.22 \\ 
LWP 26887 &9324.91875 &18.95$\pm$0.20 & &LWP 26948 &9328.14762 &20.28$\pm$0.23 \\ 
LWP 26888 &9324.96537 &18.89$\pm$0.20 & &LWP 26949 &9328.20417 &20.33$\pm$0.23 \\ 
LWP 26889 &9325.00846 &19.25$\pm$0.20 & &LWP 26950 &9328.24645 &20.45$\pm$0.23 \\ 
LWP 26890 &9325.06462 &19.07$\pm$0.20 & &LWP 26951 &9328.29373 &20.66$\pm$0.23 \\ 
LWP 26891 &9325.10877 &18.85$\pm$0.20 & &LWP 26952 &9328.33689 &20.66$\pm$0.24 \\ 
LWP 26892 &9325.15744 &18.82$\pm$0.20 & &LWP 26953 &9328.38307 &20.18$\pm$0.23 \\ 

\end{tabular}

\begin{tabular}{ccccccc}
\multicolumn{7}{c}{TABLE 3 -- {\it Continued}}\\
\\[0.2cm]
\hline
\hline
\\[0.01cm]
\multicolumn{1}{c}{Image} &
\multicolumn{1}{c}{Julian Date} &
\multicolumn{1}{c}{$F_{\lambda}(2688$\,\AA)} &{~~~} &
\multicolumn{1}{c}{Image} &
\multicolumn{1}{c}{Julian Date} &
\multicolumn{1}{c}{$F_{\lambda}(2688$\,\AA)} \\
\multicolumn{1}{c}{} & {(2,440,000+)} & & &
\multicolumn{1}{c}{} & {(2,440,000+)} & \\
\\[0.01cm]
\hline

LWP 26954 &9328.55602 &19.67$\pm$0.23 & &LWP 26992 &9330.36723 &20.14$\pm$0.22 \\ 
LWP 26955 &9328.59917 &21.06$\pm$0.23 & &LWP 26993 &9330.41095 &19.83$\pm$0.22 \\ 
LWP 26956 &9328.64231 &20.77$\pm$0.25 & &LWP 26994 &9330.54402 &19.61$\pm$0.22 \\ 
LWP 26957 &9328.68395 &20.87$\pm$0.28 & &LWP 26995 &9330.58879 &19.44$\pm$0.22 \\ 
LWP 26958 &9328.72639 &21.36$\pm$0.29 & &LWP 26996 &9330.63270 &20.07$\pm$0.22 \\ 
LWP 26959 &9328.76899 &20.26$\pm$0.23 & &LWP 26997 &9330.67968 &20.87$\pm$0.24 \\ 
LWP 26960 &9328.81123 &20.22$\pm$0.22 & &LWP 26998 &9330.72290 &21.63$\pm$0.25 \\ 
LWP 26961 &9328.87379 &20.70$\pm$0.21 & &LWP 26999 &9330.76601 &21.46$\pm$0.24 \\ 
LWP 26962 &9328.91296 &20.56$\pm$0.22 & &LWP 27000 &9330.81270 &20.24$\pm$0.23 \\ 
LWP 26963 &9328.95503 &20.49$\pm$0.22 & &LWP 27001 &9330.88433 &19.60$\pm$0.22 \\ 
LWP 26964 &9329.00142 &20.84$\pm$0.23 & &LWP 27002 &9330.93458 &20.11$\pm$0.23 \\ 
LWP 26965 &9329.04812 &21.26$\pm$0.23 & &LWP 27003 &9330.98451 &20.02$\pm$0.22 \\ 
LWP 26966 &9329.08977 &20.84$\pm$0.23 & &LWP 27004 &9331.02991 &20.42$\pm$0.23 \\ 
LWP 26967 &9329.13165 &21.19$\pm$0.23 & &LWP 27005 &9331.08154 &20.42$\pm$0.23 \\ 
LWP 26968 &9329.17823 &20.97$\pm$0.23 & &LWP 27006 &9331.13067 &20.50$\pm$0.23 \\ 
LWP 26969 &9329.23031 &20.88$\pm$0.24 & &LWP 27007 &9331.17780 &20.30$\pm$0.23 \\ 
LWP 26970 &9329.27590 &20.92$\pm$0.24 & &LWP 27009 &9331.27549 &19.88$\pm$0.22 \\ 
LWP 26971 &9329.32187 &21.00$\pm$0.24 & &LWP 27010 &9331.32125 &19.83$\pm$0.23 \\ 
LWP 26972 &9329.36632 &20.68$\pm$0.24 & &LWP 27012 &9331.40681 &19.52$\pm$0.23 \\ 
LWP 26973 &9329.41200 &20.13$\pm$0.23 & &LWP 27013 &9331.53384 &19.20$\pm$0.22 \\ 
LWP 26974 &9329.54823 &20.50$\pm$0.22 & &LWP 27014 &9331.58003 &19.96$\pm$0.22 \\ 
LWP 26975 &9329.59112 &20.43$\pm$0.22 & &LWP 27015 &9331.62443 &19.72$\pm$0.22 \\ 
LWP 26976 &9329.63353 &20.07$\pm$0.22 & &LWP 27016 &9331.66881 &20.02$\pm$0.23 \\ 
LWP 26977 &9329.67546 &19.43$\pm$0.22 & &LWP 27017 &9331.71460 &19.90$\pm$0.24 \\ 
LWP 26978 &9329.71931 &19.46$\pm$0.22 & &LWP 27019 &9331.82530 &19.44$\pm$0.23 \\ 
LWP 26979 &9329.76061 &19.53$\pm$0.22 & &LWP 27024 &9332.78837 &20.31$\pm$0.22 \\ 
LWP 26980 &9329.80323 &19.43$\pm$0.22 & &LWP 27025 &9332.82462 &20.56$\pm$0.22 \\ 
LWP 26981 &9329.86347 &19.19$\pm$0.22 & &LWP 27030 &9333.75813 &19.64$\pm$0.21 \\ 
LWP 26982 &9329.90594 &18.75$\pm$0.21 & &LWP 27031 &9333.80360 &19.82$\pm$0.19 \\ 
LWP 26983 &9329.94869 &19.03$\pm$0.21 & &LWP 27032 &9333.84854 &19.86$\pm$0.19 \\ 
LWP 26984 &9329.99855 &18.73$\pm$0.21 & &LWP 27034 &9334.76765 &19.32$\pm$0.19 \\ 
LWP 26985 &9330.04741 &19.24$\pm$0.22 & &LWP 27035 &9334.81023 &19.93$\pm$0.21 \\ 
LWP 26986 &9330.09278 &18.88$\pm$0.21 & &LWP 27036 &9334.85409 &19.29$\pm$0.20 \\ 
LWP 26987 &9330.13465 &19.05$\pm$0.22 & &LWP 27040 &9335.76131 &19.98$\pm$0.19 \\ 
LWP 26988 &9330.17782 &19.12$\pm$0.22 & &LWP 27041 &9335.80560 &20.07$\pm$0.21 \\ 
LWP 26989 &9330.22801 &19.56$\pm$0.22 & &LWP 27042 &9335.84938 &20.29$\pm$0.20 \\ 
LWP 26990 &9330.27389 &19.53$\pm$0.22 & &LWP 27048 &9336.76696 &20.01$\pm$0.20 \\ 
LWP 26991 &9330.31899 &19.72$\pm$0.22 & &LWP 27050 &9336.85257 &20.08$\pm$0.21 \\ 
\\[0.01cm]
\hline
\multicolumn{7}{l}{
$^a$Rest-frame continuum fluxes in units of 10$^{-14}$ ergs
s$^{-1}$ cm$^{-2}$ \AA$^{-1}$.}\\
\end{tabular}
\end{center}

\newpage
\small
\begin{table}[h]
\begin{center}
\begin{tabular}{ccccc}
\multicolumn{5}{c}{TABLE 4}
\\[0.2cm]
\multicolumn{5}{c}{VARIABILITY PARAMETERS}
\\[0.2cm]
\hline
\hline
\\[0.01cm]
\multicolumn{1}{c}{Feature}&
\multicolumn{1}{c}{Mean Flux$^{a}$}&
\multicolumn{1}{c}{Mean Error}&
\multicolumn{1}{c}{~F$_{var}$~}&
\multicolumn{1}{c}{~~~R$_{max}$~~~}\\
\\[0.01cm]
\hline
$F_{\lambda}(1275$\,\AA) & 43.45 & 0.009 & 0.091 & 1.51 \\
$F_{\lambda}(1440$\,\AA) & 37.66 & 0.015 & 0.070 & 1.45 \\
$F_{\lambda}(1820$\,\AA) & 28.83 & 0.013 & 0.052 & 1.31 \\
$F_{\lambda}(2688$\,\AA) & 19.83 & 0.011 & 0.042 & 1.24 \\
 & & & & \\
F(\civ)                  & 45.95 & 0.014 & 0.054 & 1.33 \\
F(\heii)                 & ~6.54 & 0.042 & 0.077 & 1.70 \\
F(\ciii])                & ~6.50 & 0.034 & 0.077 & 1.49 \\
\\[0.01cm]
\hline
\multicolumn{5}{l}{
$^a$Continuum fluxes in units of 10$^{-14}$ ergs
s$^{-1}$ cm$^{-2}$ \AA$^{-1}$.}\\
\multicolumn{5}{l}{
Line fluxes in units of
10$^{-12}$ ergs s$^{-1}$ cm$^{-2}$.}\\
\end{tabular}
\end{center}
\end{table}

\newpage
\normalsize
\begin{center}
{\bf REFERENCES}
\end{center}
\bigskip

\leftskip=0.2in
\parindent=-0.2in

Ayres, T. 1993, PASP, 105, 538

Bevington, P. R. 1969, Data Reduction and Error Analysis
for the Physical Sciences, 200

Bromage, G.E., et al. 1985, MNRAS, 215, 1

Burstein, D., \& Heiles, C. 1982, AJ, 87, 1165

Carini, M., \& Weinstein, D. 1992, NASA IUE Newsletter, 49, 5

Clavel, J. et al. 1987, ApJ, 321, 251

Clavel, J., et al. 1990, MNRAS, 246, 668

Clavel, J., et al. 1991, ApJ, 366, 64

Clavel, J. et al. 1992, ApJ, 393, 113

Collin-Souffrin, S. 1991, A\&A, 249, 344

Courvoisier, T.J.-L. \& Clavel, J. 1991, A\&A, 248, 349

Dietrich, M., et al. 1993, ApJ, 408, 416

Edelson, R.A., \& Krolik, J.H. 1988, ApJ, 333, 646

Edelson, R.A., et al. 1995, ApJ, 438, 120.

Edelson, R.A., et al. 1996, in preparation (Paper IV)

Evans, I.N., et al. 1993, ApJ, 417, 82

Ferland, G.J., \&\ Mushotzky, R.F. 1982, ApJ, 262, 564

Gaskell, C.M., Koratkar, A.P., \& Sparke, L.S. 1988, in Active
Galactic Nuclei, ed. D.E. Osterbrock and J.S. Miller, Dordrecht:
Kluwer Academic Pub., 93 

Gaskell, C.M., \& Peterson, B.M., 1987, ApJS, 65, 1

Gaskell, C.M., \& Sparke, L.S. 1986, ApJ, 305, 175

Holt, S.S., et al. 1980, ApJ, 241, L13

James, F., \& Roos M. 1975, Computer Physics
Communications,  10, 343 

Johnston, K.J., Elvis, M., Kjer, D., \& Shen, B.S.P. 1982, ApJ, 262, 61 

Kaspi, S., et al. 1996, in preparation (paper II)

Kinney, A.L., Bohlin, R.C., \& Neill, J.D. 1991 PASP, 103, 694

Korista, K.T., et al. 1995, ApJS, 97 285

Kriss, G. A., et al. 1992, ApJ, 392, 485

Krolik, J.H., et al. 1991, ApJ, 371, 541

Maoz, D., et al. 1991, ApJ, 367, 493

Osterbrock, D.E. \& Koski, A.T. 1976, MNRAS, 176, 61P

Penston, M.V., et al. 1981, MNRAS, 196, 857

Penton, S.V., et al. 1996, in preparation

Perola, G.C., et al. 1982, MNRAS 200, 293

Peterson, B.M., 1993, PASP, 105, 247

Peterson, B.M. \& Cota, S.A. 1988, ApJ, 330, 111

Peterson, B.M., et al. 1991, ApJ, 368, 119

Peterson, B.M., et al. 1992, ApJ, 392, 470

Peterson, B.M., et al. 1994, ApJ, 425, 622

Reichert, G.A., et al. 1994, ApJ, 425, 582

Savage, B.D., \&\ Mathis, J.S. 1979, ARA\&A, 17, 73

Shakura, R.I. \& Sunyaev, R.A. 1973, A\&A, 24, 337

Simkin, S.M. 1975, ApJ, 200, 567

Stark, A.A., et al. 1992, ApJS, 79, 77

Stirpe, G.M., et al. 1994, ApJ, 425, 609

Ulrich, M.-H., et al. 1985, Nature, 313, 745

Ulrich, M.-H., et al. 1991, ApJ, 382, 483

Warwick, R.S., et al. 1996, in preparation (paper III)

Weaver, K. A., et al. 1994a, ApJ, 423, 621

Weaver, K., Yaqoob, T., Holt, S. S., Mushotzky, R. F., 
Matsuoka, M., \& Yamauchi, M. 1994b, ApJ, 436, L27

White, R.J., \& Peterson, B.M. 1994, ApJ, 106, 879

Wilson, A.S., \& Ulvestad, J.S. 1982, ApJ 263, 576

Yaqoob, T., Warwick, R., \& Pounds, K. 1989, MNRAS, 236, 153

\leftskip=0.in
\parindent=0.0in
\normalsize
\newpage
\begin{center}
{\bf FIGURE CAPTIONS}
\end{center}
\bigskip
\par
{\sc FIG. 1}-- Averaged and combined SWP and LWP spectrum of NGC
4151 for the 1993 campaign. The 1975 -- 2500 \AA\ region has been 
smoothed with a 10 pixel ($\sim$15 \AA) boxcar filter for display 
purposes.

\bigskip

{\sc FIG. 2}-- Sample spectrum (SWP 49555) and spectral fits. The solid 
line is the observed spectrum, the dotted lines are the spectral 
components, and the dotted/dashed line is the sum of the components.

\bigskip

{\sc FIG. 3}-- Continuum fluxes in the 1275 \AA\ band. Fluxes from
TOMSIPS are plotted as a function of those from IUESIPS in units of
10$^{-14}$ ergs s$^{-1}$ cm$^{-2}$ \AA$^{-1}$. 

\bigskip

{\sc FIG. 4}-- IUE continuum fluxes in units of 10$^{-14}$ ergs
s$^{-1}$ cm$^{-2}$ \AA$^{-1}$ are plotted as a function of Julian
Date. The fluxes are at the midpoints of the error bars
($\pm$1$\sigma$). 

\bigskip

{\sc FIG. 5}-- IUE line fluxes in units of 10$^{-12}$ ergs s$^{-1}$
cm$^{-2}$ are plotted as a function of Julian Date. The fluxes are at
the midpoints of the error bars ($\pm$1$\sigma$). The continuum light 
curve at 1275 \AA\ is repeated in the top panel for comparison.

\bigskip

{\sc FIG. 6}-- Cross-correlation of the \civ\ line with itself (ACF)
and cross-correlation of the 1275 \AA\ continuum band with the emission
lines for the continuous data set. The CCF is given by the smooth
curve, and the DCF is given by the plotted points and error bars. 

\bigskip

{\sc FIG. 7}-- Cross-correlation of the \civ\ line with itself (ACF)
and cross-correlation of the 1275 \AA\ continuum band with the emission
lines for the entire data set. The CCF is given by the smooth
curve, and the DCF is given by the plotted points and error bars. 

\bigskip

\pagestyle{empty}

\newpage
\centerline{\psfig{figure=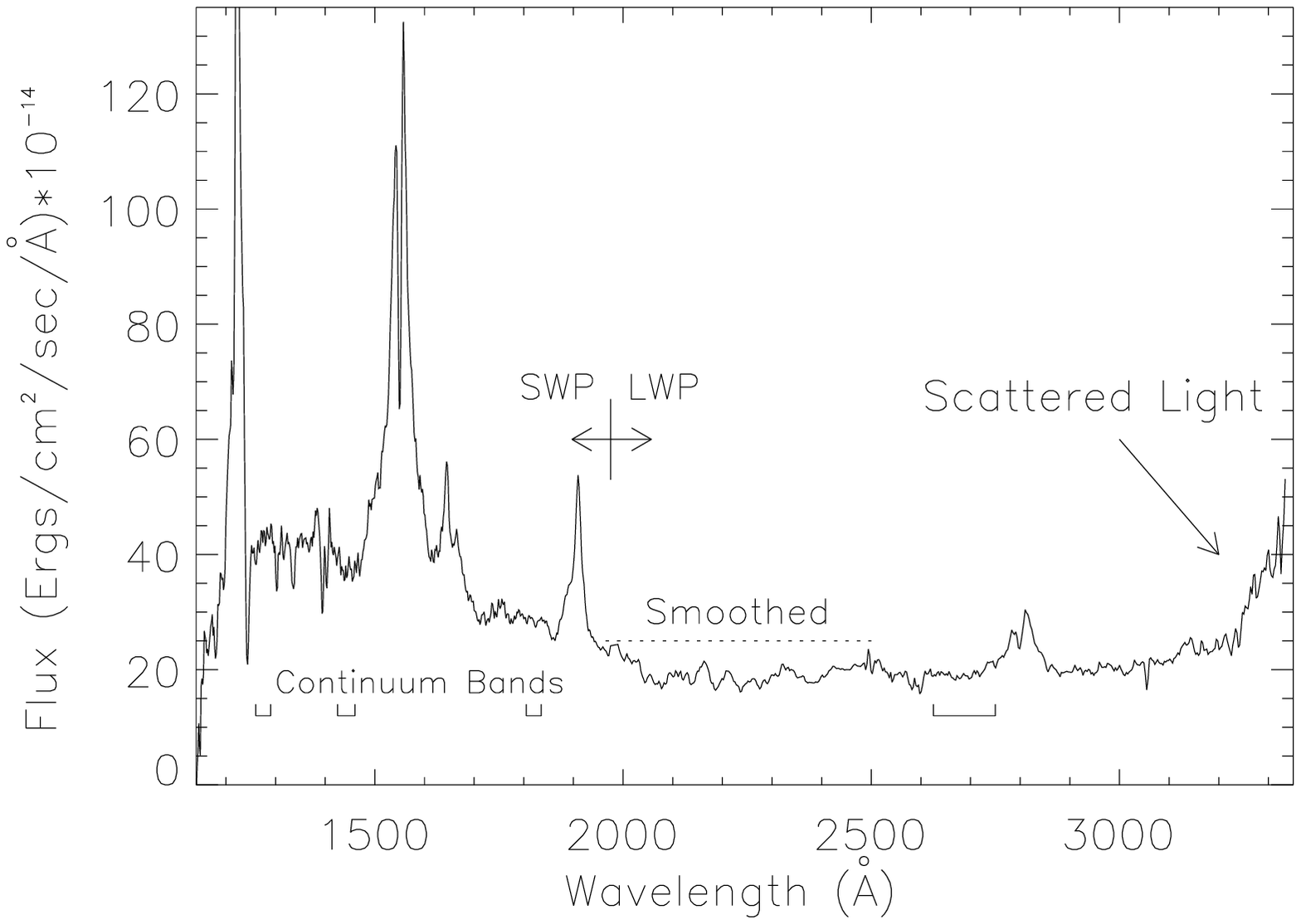}}
\bigskip
\centerline{Figure 1.}

\newpage
\centerline{\psfig{figure=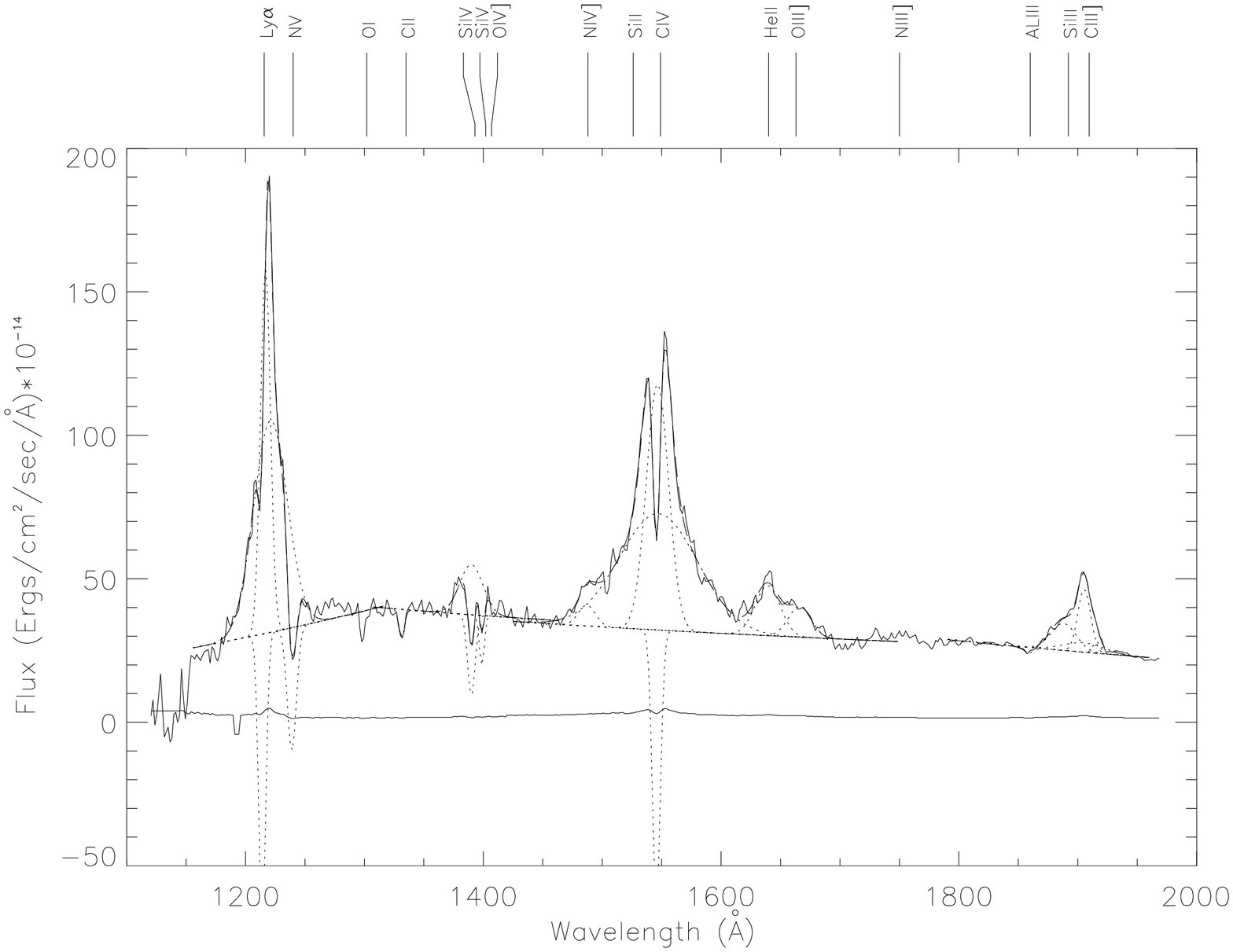,height=4.5in,width=6.5in,angle=90}}
\bigskip
\centerline{Figure 2.}

\newpage
\centerline{\psfig{figure=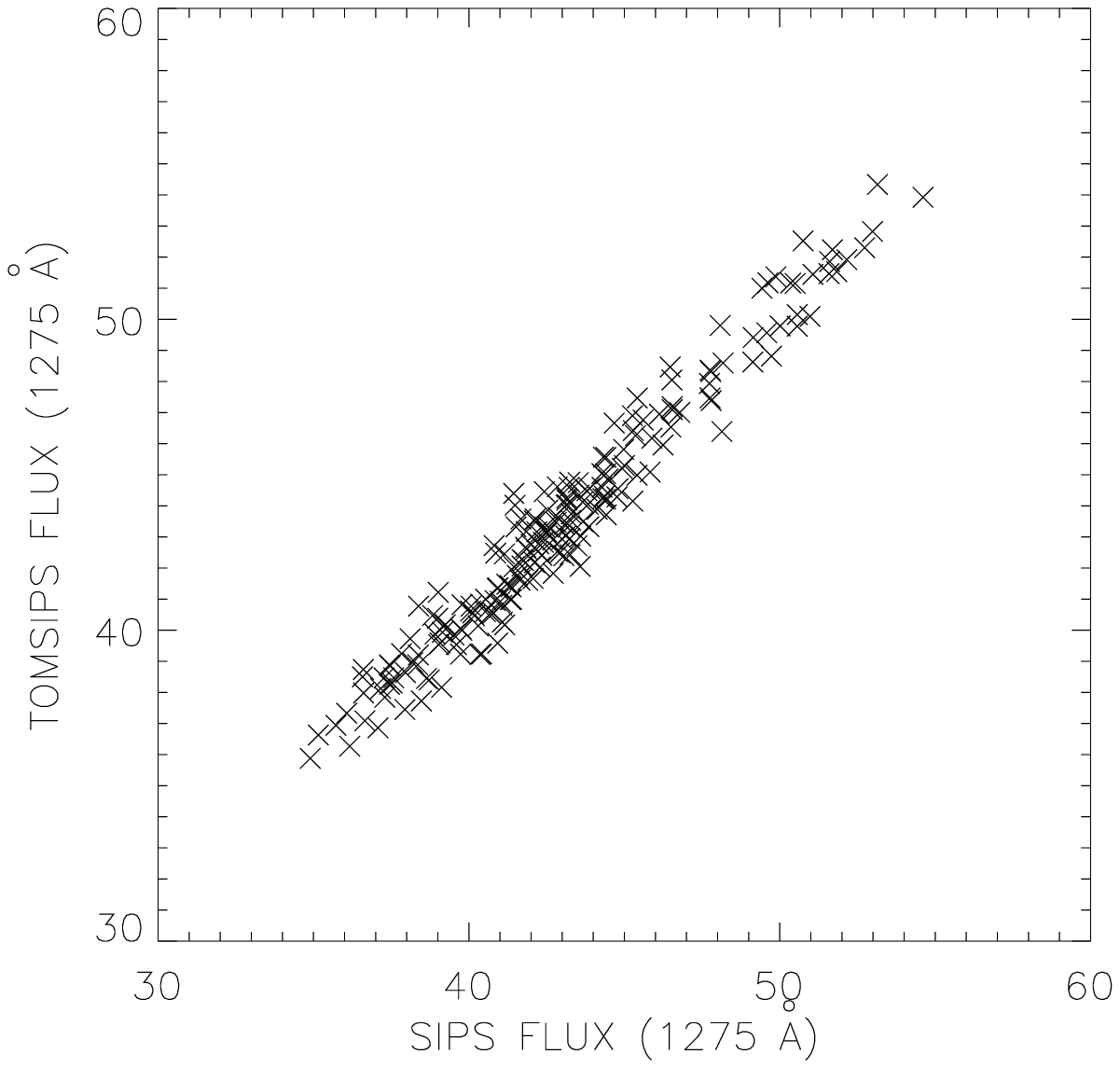}}
\bigskip
\centerline{Figure 3.}

\newpage
\centerline{\psfig{figure=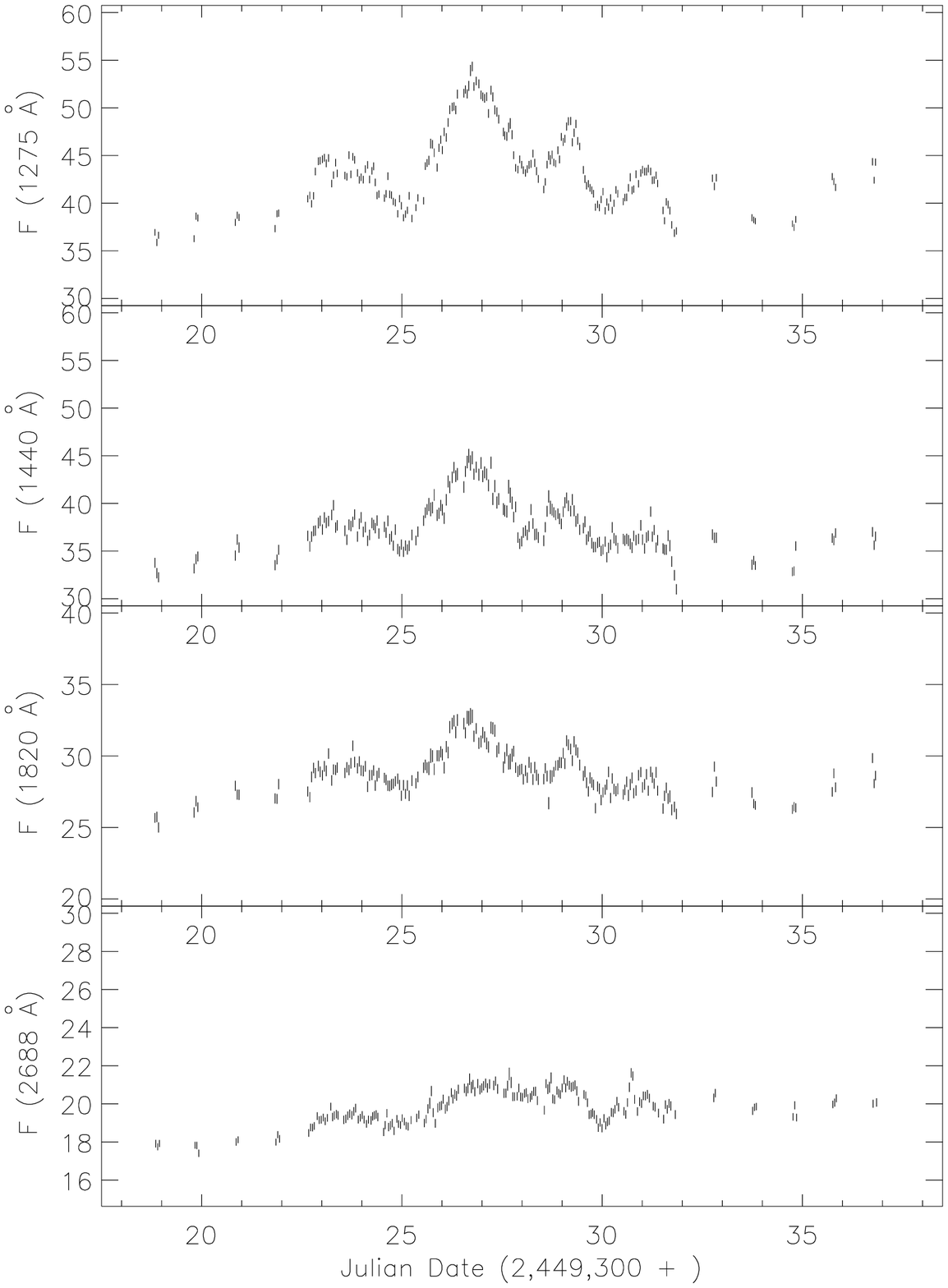,height=8in,width=6in}}
\bigskip
\centerline{Figure 4.}

\newpage
\centerline{\psfig{figure=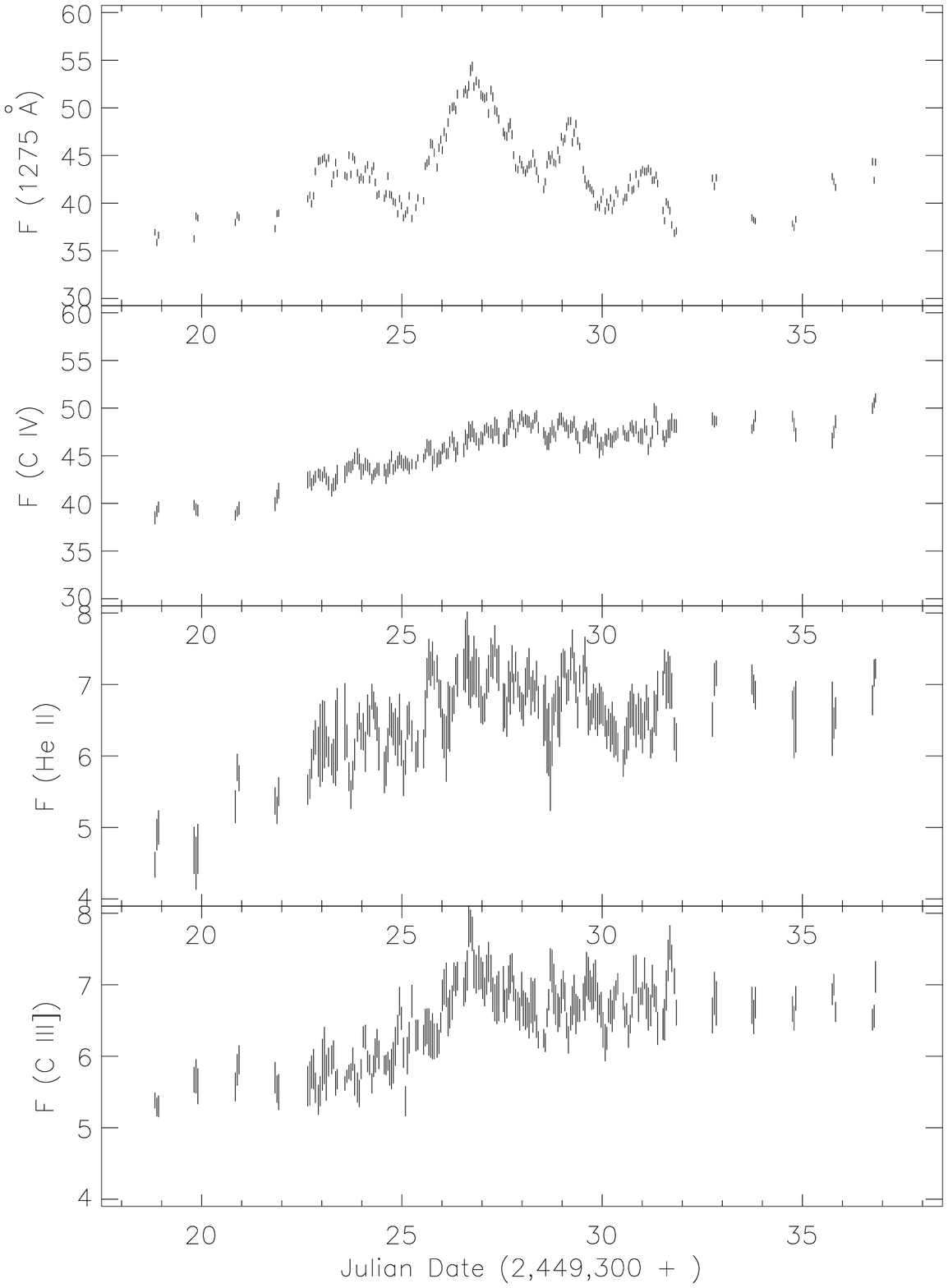,height=8in,width=6in}}
\bigskip
\centerline{Figure 5.}

\newpage
\centerline{\psfig{figure=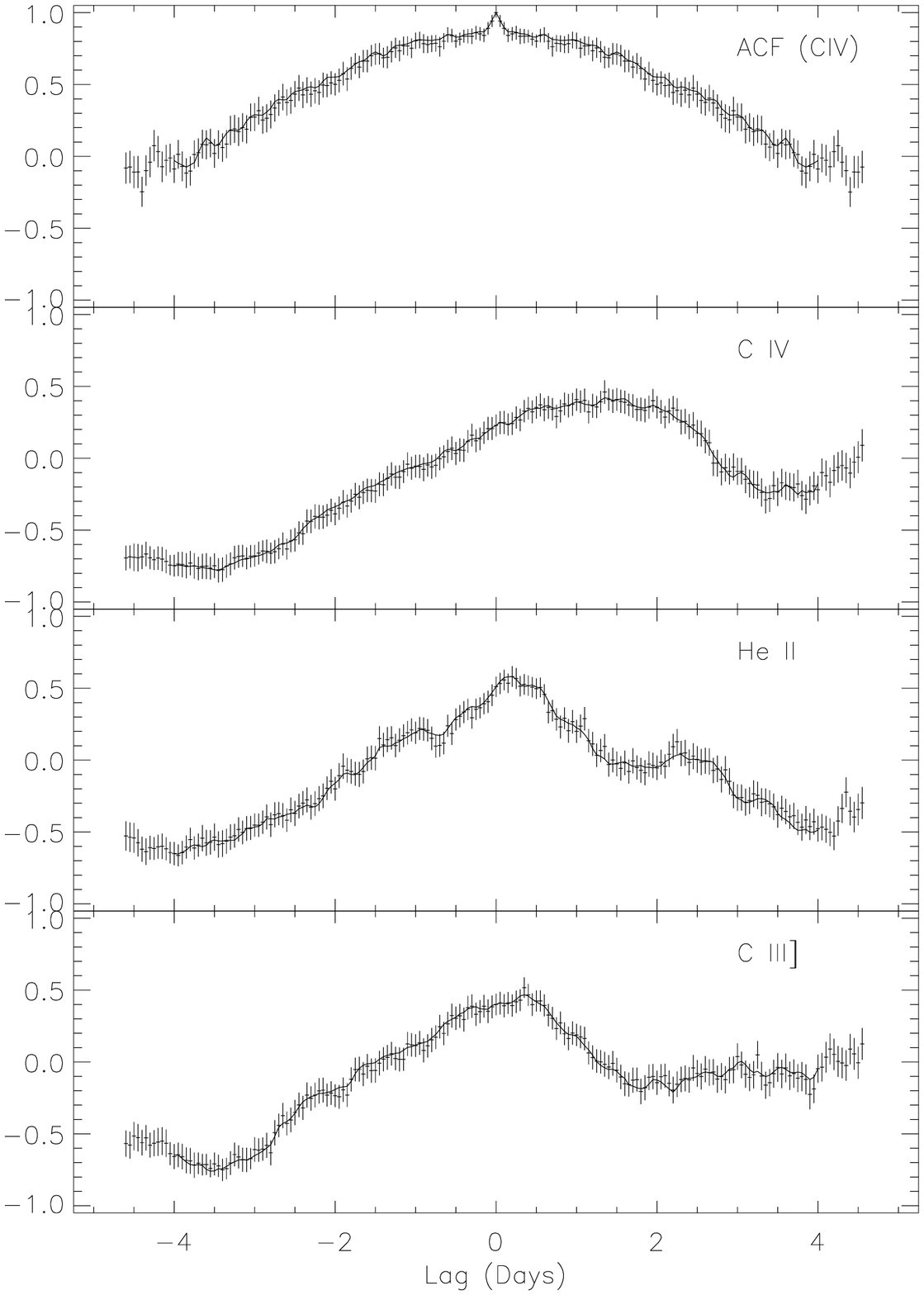,height=8in,width=6in}}
\bigskip
\centerline{Figure 6.}

\newpage
\centerline{\psfig{figure=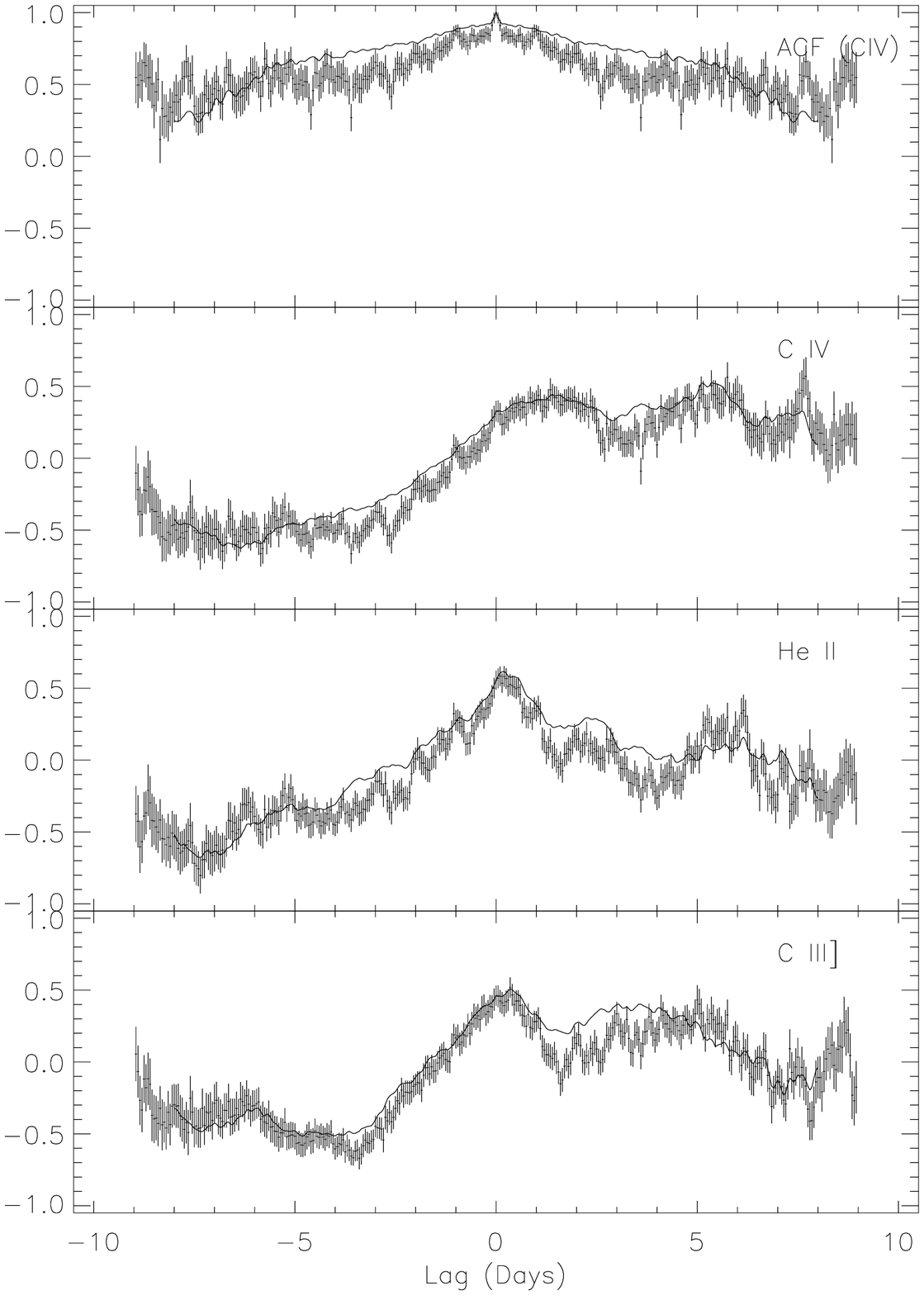,height=8in,width=6in}}
\bigskip
\centerline{Figure 7.}

\end{document}